\documentclass[modern]{aastex63}
\usepackage{comment}
\usepackage{xspace}

\newcommand{\uniform}{{uniform}\xspace}
\newcommand{\loguniform}{{log-uniform}\xspace}
\newcommand{\Aul}{$A^{95\%}_{\rm GWB}$\xspace}
\newcommand{\Agwb}{$A_{\rm GWB}$\xspace}
\newcommand{\Arn}{$A_{\rm RN}$\xspace}

\defcitealias{abb+18a}{NG11}
\defcitealias{abb+18b}{NG11gwb}

\revised{\today}
\submitjournal{ApJL}

\shorttitle{Model Dependent GWB Statistics}
\shortauthors{Hazboun et al.}

\graphicspath{{./}{figures/}}

\begin{document}

\title{Model Dependence of Bayesian Gravitational-Wave Background Statistics 
for Pulsar Timing Arrays}

\correspondingauthor{Jeffrey S. Hazboun}
\email{hazboun@uw.edu}

\author[0000-0003-2742-3321]{Jeffrey S. Hazboun}
\affiliation{Physical Sciences Division, University of Washington Bothell, 18115 Campus Way NE, Bothell, WA 98011, USA}
\author[0000-0003-1407-6607]{Joseph Simon}
\affiliation{Jet Propulsion Laboratory, California Institute of Technology, 4800 Oak Grove Drive, Pasadena, CA 91109, USA}
\affiliation{Department of Astrophysical and Planetary Sciences, University of Colorado, Boulder, CO 80309, USA}
\author[0000-0002-7778-2990]{Xavier Siemens}
\affiliation{Department of Physics, Oregon State University, Corvallis, OR 97331, USA}
\author[0000-0003-4915-3246]{Joseph D.~Romano}
\affiliation{Department of Physics and Astronomy, Texas Tech
University, Lubbock, TX 79409-1051, USA}

\begin{abstract}

Pulsar timing array (PTA) searches for a gravitational-wave background 
(GWB) typically include time-correlated ``red'' noise models intrinsic to
each pulsar. Using a simple simulated PTA dataset with an injected GWB
signal we show that the details of the red noise models used,
including the choice of amplitude priors and even which pulsars
{\it have} red noise, have a striking impact on the GWB statistics, 
including both upper limits and estimates of the GWB amplitude. We
find that the standard use of \uniform priors on the red
noise amplitude leads to $95\%$ upper limits, as calculated from
one-sided Bayesian credible intervals, that are less than the injected
GWB amplitude $50\%$ of the time. In addition, amplitude estimates of
the GWB are systematically {\it lower} than the injected value by 
$10-40\%$, depending on which models
and priors are chosen for the intrinsic red noise. We tally the
effects of model and prior choice and demonstrate how a ``dropout''
model, which allows flexible use of red noise models in a Bayesian
approach, can improve GWB estimates throughout.

\end{abstract}

\keywords{pulsar timing arrays, gravitational waves, multimessenger astronomy, data analysis}

\section{Introduction} 
\label{sec:intro}

Pulsar timing arrays (PTAs) are sensitive to gravitational waves (GWs) in the nanohertz frequency band \citep{saz78, det79,fb90}.
The most promising GW sources in that band are supermassive binary
black holes (SMBBHs) that are formed via mergers of massive galaxies
\citep{rsg2015}. Orbiting SMBBHs can produce a gravitational-wave stochastic background (GWB), individual periodic signals, and transient GW bursts \citep{spolaor:2018}. 
The GWB from SMBBHs manifests in pulsar timing data as a stochastic
signal that is correlated both temporally and spatially between pulsars
\citep{hd83,p01,jb03}. The spatial correlations are defined by the overlap
reduction function known as the Hellings-Downs curve~\citep{hd83}, and
the temporal correlations follow a steep power-law
spectrum~\citep{p01,jb03}---i.e., the GWB is a spatially-correlated ``red''
noise process in pulsar timing data. The spatial correlations are a
direct consequence of general relativity, originating from the
quadrupolar nature of GWs.

However, there are also a number of potential non-GW sources of 
temporal correlations (red noise processes) in pulsar timing data.
Intrinsic spin noise exists in many canonical
pulsars~\citep{cordes1985} and at a lower level in millisecond
pulsars~\citep{cordes2013}. Clock errors and solar system ephemeris
errors can manifest as spatially-correlated sources of red
noise~\citep{champion2010,thk+2016}. Lastly, unmodeled trends in
radio-dependent propagation delays, e.g., dispersion and scattering due to the interstellar medium, will also produce red noise in pulsar timing data \citep{cordes2010}. Since the stochastic GWB is modeled as a red noise process with a steep spectral index, unmodeled or mismodeled noise in pulsars can have an impact on GWB detection statistics and parameter estimation~\citep{hazboun:2020slice}.  It is therefore imperative to accurately model the noise in individual pulsars so that it does not contaminate the GWB signal. 

As PTA datasets have matured and reached astrophysically interesting
levels of sensitivity, $95\%$ upper limits (ULs) on the amplitude of 
the GWB, \Aul, have been the flagship statistic quoted by PTA collaborations
\citep{srl+15,Lentati:2016ygu,abb+16,abb+18b}.
These ULs have been used to constrain model parameters for SMBBH populations \citep{sbs16, abb+16}, such as the $M-M_\mathrm{bulge}$ relationship between SMBHs to their host galaxies, and to calculate the evidence for various SMBBH population models \citep{srl+15,shk+18}.
The standard \Aul quoted is the one-sided credible interval of a Bayesian
analysis and is therefore subject to the choice of the signal+noise 
models, including the choice of prior probability distributions for 
the parameters associated with these models.

\subsection{Statement of the problem}
\label{sec:problem}

In this paper, we systematically investigate how the choice of 
different signal+noise models, including the choice of 
priors, affects the estimates and ULs of the GWB returned by
PTA analyses.
We shall see that of utmost importance is the choice of noise 
model for the individual pulsars\footnote{As another pertinent example of
the interplay of noise models and GWs, the development of the ECORR noise 
parameter \citep{abb+14} was in response to spurious
single-source GW detections at frequencies higher than $1/{\rm yr}$
caused by noise correlated across frequencies on intraday timescales.}, as steep red noise in 
one or several pulsars can masquerade as red noise in the GWB,
potentially leading to a bias in the conditional median estimate of the
amplitude of the background, \Agwb, or its UL, \Aul.
Not surprisingly, a marginally significant GWB can be 
absorbed by red noise models for individual pulsars across 
the PTA, reducing the estimated amplitude of the GWB.
It is also possible for models that do not accurately model
the pulsar noise to lead to spurious increases in the estimated 
GWB amplitude. 
As shown in \citet{hazboun:2020slice},
the significance of the detection of a GWB signal is strongly 
affected by which red noise models are chosen for each pulsar.

Over the years the above considerations regarding red noise 
led the PTA community to adopt a conservative approach that includes a red noise model for {\em every} pulsar in the array
\citep{yardley2011,dfg+13,lentati2015,Lentati:2016ygu,abb+16,abb+18b}.
An obvious alternative to this approach is to not model 
intrinsic pulsar red noise at all, which would effectively attribute any 
observed red noise to the GWB.
This approach is arguably
{\em more} conservative than the standard approach, since it results in the largest ULs on the GWB amplitude. 
A third choice is to allow a Bayesian analysis to {\em choose} 
which pulsars should include intrinsic red noise models---i.e., 
letting the data decide whether a red noise model for a
particular pulsar is preferred to a white-noise-only model in a search for the GWB.
There are a number of possibilities for implementing this more
flexible pulsar noise model, including trans-dimensional models
\citep{ellis:2016}, hierarchical modeling \citep{gelman:2007} and
product space methods \citep{taylor2020,cc95,g01,hee15}.
But for the analyses that we perform in this paper, 
we adopt a {\em dropout} method \citep{aab+19} 
(discussed in more detail in \autoref{sec:dropout}) to decide 
which pulsars should be assigned red noise.
{\em As we shall see below, a pulsar red noise model that utilizes 
the dropout method, and allows the data to decide whether or not to include intrinsic red noise in each pulsar, gives results that most robustly and accurately return the injected \Agwb.}

\subsection{Outline of the paper}
\label{sec:outline}

To compare the effects of different pulsar noise models on the 
statistics of the GWB, we analyze 400 
realizations of simulated PTA data consisting of a  GWB 
signal injected into white timing noise (WN).
Details of the simulations, signal+noise models, and data analysis 
methods used are discussed in detail in \autoref{sec:sims}.
We allow for different priors for both the amplitude of the 
GWB and the red noise of the individual pulsars, 
as well as whether or not a red noise 
model for a particular pulsar should be included, i.e., the dropout method.
The results are described in \autoref{sec:results} for both GWB
parameter estimation (\autoref{sec:param_estim}) and 95\% UL
calculations (\autoref{sec:ul_skew}).
It turns out that the choice of the individual red noise models 
has a surprisingly strong effect, especially in the case of UL analyses.
We consider more realistic simulations in \autoref{sec:do_rn},
where we inject red timing noise for a handful of pulsars, and 
show that the dropout method can also handle this more realistic
scenario without any problems.
Finally, in \autoref{sec:11yr}, we reanalyze the NANOGrav 11-year
dataset using the dropout method, obtaining a revised 95\% UL,
$A_{\rm GWB}^{95\%} = 3.0\times 10^{-15}$.  This is more than 
twice as large as the value reported in \citet{abb+18b}.

%%%%%%%%%%%%%%%%%%%%%%%%%%%%%%%%%%%%%%%%%%%%%%%%%%%%%%%%%%
\section{Simulated Data and Signal+Noise Models}
\label{sec:sims}

To investigate the effect of different models and priors for the intrinsic red
noise in pulsars, we performed a number of simulations and analyses. The software {\tt libstempo} \citep{libstempo} is used to simulate
400 realizations of a GWB with amplitude \Agwb$=1.4\times10^{-15}$ in
WN at a level of $1\,\mu s$ for a simple PTA 
dataset based on the pulsars in the IPTA's second mock data challenge
\citep{hazboun:2018mdc}. 
The WN is simulated using the time of arrival (TOA) errors, identical in amplitude 
for all pulsars. (Pulse TOAs are the fiducial data used in PTA analyses.)
The amplitude of the WN is treated as a known quantity for all of 
the analyses.
We then compare the results of the different models and priors to the results of a model that incorporates only what we know to be in the simulated data: a red GWB signal plus WN. 

The construction of the likelihood and analysis methods match 
exactly those used in recent PTA data analysis work such as
\citet{abb+18a} and developed over the last decade in the literature
\citep{tlb+17,abb+16,vHv:2014,abb+14,taylor2013,esvh13}.
Therefore, we do not describe the analysis methods in detail here except to note a few features connected to the signal+noise models relevant to this work. 

The GWB is modeled as a Gaussian process \citep{rw06} in the 
Fourier domain \citep{vHv:2014, Lentati:2016ygu} with a 
power spectral density given by a power-law.
The main results quoted from searches for the GWB assume a {\em
fixed} spectral index $\gamma=13/3$ for the induced timing residuals:
\begin{equation}
P_g(f) = \frac{A_{\rm GWB}^2}{12\pi^2} \left(\frac{f}{f_{\rm
yr}}\right)^{-\gamma}{f_{\rm yr}}^{-3}\,,
\end{equation}
where $f_{\rm yr}\equiv 1/{\rm year}$ is a reference frequency.
The choice $\gamma=13/3$ corresponds to a spectral index of $-2/3$ 
for the characteristic strain of the GWB, appropriate for 
inspiraling binaries \citep{p01,jb03}. 
Typical analyses, in addition to the GWB, include a separate red noise
model for each pulsar, parametrized in the same way as the GWB, 
\begin{equation}
P_{\rm RN}(f) = \frac{A_{\rm RN}^2}{12\pi^2} \left(\frac{f}{f_{\rm
yr}}\right)^{-\gamma_{\rm RN}}{f_{\rm yr}}^{-3}\,,
\end{equation}
where {\em both} the spectral index $\gamma_{\rm RN}$ and red 
noise amplitude $A_{\rm RN}$ are 
allowed to vary for each pulsar. This adds $2N_{\rm pulsars}$
parameters to the search. The prior on the spectral index is taken to
be uniform from $0$ to $7$. This covers the range from white noise 
($\gamma=0$) to the
the steepest power spectral density for which the quadratic spin down removes
dependence on any lower cutoff frequency in that power spectral density 
\citep{vanHaasteren:2012hj, blandford+1984}. 
In principle, the prior on the spectral index could also be chosen 
differently (see, e.g., \citet{Callister:2017}), but we do not 
investigate the effects of those choices here.
However, we do consider the effect of different priors on the 
{\em amplitudes} of both the GWB and pulsar red noise, \Agwb and \Arn. 
We use either {\em uniform} or {\em log-uniform} (uniform in log space) probability 
distributions for these individual amplitudes, defined over the 
range of values $10^{-18}$ to $10^{-14}$.
\autoref{tab:models} lists the various signal+noise models and prior
probability distributions used in our analyses. Note that in most cases the GW signal does not include Hellings-Downs spatial correlations as these considerably decrease the computational efficiency. The alternative signal model searches for a common red noise process \citep{abb+18b} are subscripted in \autoref{tab:models} with {\sc crn}. 

\begin{deluxetable*}{l|l|l}
\tablenum{1}
\tablecaption{Different signal+noise models and prior 
probability distributions used in our analyses:
RN stands for the pulsar red noise model;
DO stands for the red noise dropout model;
CRN stands for a signal model without spatial correlations; and
HD stands for a signal model that includes spatial Hellings-Downs
correlations. 
For the pulsar red noise models, there are different 
amplitude and spectral index parameters for {\em each} pulsar, for a total of $2 N_{\rm pulsars}$ parameters.
\label{tab:models}}
\tablewidth{0pt}
\tablehead{
\colhead{Model} & \colhead{Signal prior} & \colhead{Noise priors}}
\startdata
GWB$_{\text{\sc crn}}$-only  & 
$\pi(A_{\rm GWB})={\rm logunif}(10^{-18}, 10^{-12})$ & 
--- \\
\hline
GWB$_{\text{\sc crn}}$-only &
$\pi(A_{\rm GWB})={\rm unif}(10^{-18}, 10^{-12})$ & 
--- \\
\hline
GWB$_{\text{\sc crn}}$+RN  & 
$\pi(A_{\rm GWB})={\rm logunif}(10^{-18}, 10^{-12})$ & 
$\pi(A_{\rm RN})={\rm logunif}(10^{-20}, 10^{-11})$ 
\\
& & 
$\pi(\gamma_{\rm RN})={\rm unif}(0,7)$ 
\\
\hline
GWB$_{\text{\sc crn}}$+RN  & 
$\pi(A_{\rm GWB})= {\rm unif}(10^{-18}, 10^{-12})$ & 
$\pi(A_{\rm RN})= {\rm unif}(10^{-20}, 10^{-11})$ 
\\
& & 
$\pi(\gamma_{\rm RN})={\rm unif}(0,7)$ 
\\
\hline
GWB$_{\text{\sc crn}}$+RN$_{\text{\sc do}}$ & 
$\pi(A_{\rm GWB})={\rm logunif}(10^{-18}, 10^{-12})$ & 
$\pi(A_{\rm RN})={\rm logunif}(10^{-20}, 10^{-11})$ 
\\
& & 
$\pi(\gamma_{\rm RN})={\rm unif}(0,7)$ 
\\
\hline
GWB$_{\text{\sc crn}}$+RN$_{\text{\sc do}}$  & 
$\pi(A_{\rm GWB})= {\rm unif}(10^{-18}, 10^{-12})$ & 
$\pi(A_{\rm RN})= {\rm unif}(10^{-20}, 10^{-11})$ 
\\
& & 
$\pi(\gamma_{\rm RN})={\rm unif}(0,7)$ 
\\
\hline
GWB$_{\text{\sc hd}}$+RN & 
$\pi(A_{\rm GWB})={\rm logunif}(10^{-18}, 10^{-12})$ & 
$\pi(A_{\rm RN})={\rm logunif}(10^{-20}, 10^{-11})$ 
\\
& & 
$\pi(\gamma_{\rm RN})={\rm unif}(0,7)$ 
\\
\enddata
%\tablecomments{}
\end{deluxetable*}

All searches use the software {\sc Enterprise} \citep{enterprise} and {\tt
enterprise\_extensions} for modeling the PTA data likelihood, the GWB,
and the various signal+noise models. 
We used the Parallel Tempering Markov-Chain Monte
Carlo sampler {\tt PTMCMCSampler} \citep{ptmcmc} for sampling the
likelihood.

%%%%%%%%%%%%%%%%%%%%%%%%%%%%%%%%
\subsection{Pulsar red noise dropout method}
\label{sec:dropout}

In addition to the two models where we only search for the GWB (GWB-only in \autoref{tab:models})
or where we model intrinsic red noise for {\em every} pulsar (GWB+RN in \autoref{tab:models}),
we consider a more flexible per-pulsar noise model 
that uses the data to determine whether or not an individual pulsar
should be modeled as having red noise (GWB+RN$_{\text{\sc do}}$ in \autoref{tab:models}).
This model is implemented using the so-called
{\em dropout} method (Vigeland, in prep)
on the red noise model, which uses a discrete parameter to switch the red noise
model for a particular pulsar on or off during the 
Markov Chain Monte Carlo (MCMC) sampling 
of a Bayesian analysis. 
This extremely flexible tool has been used for investigating
the support of deterministic and stochastic signals in particular
pulsars \citep{aab+19,aab+20}. 
The GWB+RN$_{\text{\sc do}}$ analyses therefore include a red noise model for each pulsar with an amplitude and spectral index along with a dropout parameter. 
If the dropout parameter samples above a certain threshold, 
then the red noise model acts as usual. 
If the threshold is not met, then the red noise model is turned off 
completely. The threshold defines the prior odds ratio for a red noise
model to be turned on.

Throughout this work we use a threshold of $10/11$ for the red noise
model to be turned on. This means that given no support for red noise,
the red noise model will be turned on only $1/11$th of the time. 
This threshold effectively set an odds ratio of $10:1$  as a hurdle to overcome 
in order to use a red noise model for a particular pulsar.
Although one can consider using different threshold values for the 
dropout model, we do not investigate here how these different 
values affect the GWB statistics.

%%%%%%%%%%%%%%%%%%%%%%%%%%%%%%%%%%%%
\subsection{Sensitivity to choice of priors}
\label{sec:coverage}

As discussed in \autoref{sec:problem}, the statistics derived from 
Bayesian analyses depend on the
choice of signal+noise model, including the choice of prior 
probability distributions for the parameters associated with 
those models \citep{kass+1996}. 
Given sufficiently informative
data the prior choices will matter little;
however, PTA datasets are not yet at this point \citep{abb+18b}. 
As the datasets continue to increase in duration and sensitivity increases, it is critical to understand any potential pitfalls and limitations of our analysis.

Specifically,
the simulations and analyses chosen for this work 
allow us to compare the performance of different models, and their fidelity in returning injected GWB parameters, when applied to a relatively simple dataset.
If the signal+noise model matches that used in the simulations,
we should obtain, on average,
the expected coverage for our credible intervals.
For signal+noise models that do not match the simulations, 
over- or under-coverage is possible.
Thus, the analyses that we perform here can be thought of as a 
``sensitivity analysis" \citep{efron:2015}, which checks the
robustness of our statistical inference results to the choice 
of models and priors. However, these simulations do
not completely test the ``coverage'' of different signal+noise models. 
Coverage is the fidelity of Bayesian credible intervals 
(or likewise frequentist confidence intervals) over many iterations 
of an analysis \citep{heinrich:2004}, which allows us to answer the 
question ``does the injected value of a parameter fall within an 
$X\%$ interval in $X\%$ of simulations?'' In order to formally check
the coverage of a Bayesian pipeline, one would need to sample from
the prior on \Agwb, as well as look at different realizations, something
that would be too prohibitive to do across all of the models considered here.
As we shall see below, choosing the wrong model---in this case,
whether or not some or all pulsars have red noise---can 
skew the final statistics.

%%%%%%%%%%%%%%%%%%%%%%%%%%%%%%%%%%%%%%%%%%%%%%%%
\section{Results of analyses on simulated data}
\label{sec:results}

The following two subsections describe how the choice of 
different signal+noise models and prior probability distributions
affects both GWB parameter estimation and UL calculations.

\subsection{Effect of signal+noise models on GWB parameter estimation}
\label{sec:param_estim}

We begin by showing the effects of different signal+noise 
models and priors on estimates of the amplitude \Agwb of the GWB.
The first analysis that we performed, which serves as the base model 
for all comparisons, uses the GWB-only signal+noise model---i.e., 
we only search for the GWB and use TOAs weighted by their WN errors. 
Not surprisingly, given that we are analyzing the data using the 
same model that we used to produce the simulations, 
we recover posterior distributions for
the median value of $A_{\rm GWB}$ that agree well with the 
injected value, $A_{\rm GWB}^{\rm inj} = 1.4\times 10^{-15}$
(see the blue violin plots in the two panels of 
Figure~\ref{fig:median_compare}.)
The two panels correspond to two different prior 
probability distributions for $A_{\rm GWB}$, either log-uniform 
or uniform over the range $10^{-18}$ to $10^{-14}$.
Analyses using log-uniform priors are usually referred to as
``detection runs'' in the PTA literature as they are especially
effective for obtaining a Savage-Dickey approximation
\citep{dickey1971} to the Bayes factor for weak signals. 
Uniform priors on the amplitude of the GWB are usually used in 
``upper limit" runs, in order to provide more conservative ULs
on $A_{\rm GWB}$.

The other violin plots in Figure~\ref{fig:median_compare}
show the recovered 
distributions of the median value of $A_{\rm GWB}$
at two different observation times (11.5 and 15~years) 
for the 
other signal+noise models and priors listed in \autoref{tab:models}.
From these recovered distributions we are able to draw 
several conclusions:

\begin{figure}[ht!]
\begin{center}
\includegraphics[width=0.49\textwidth]{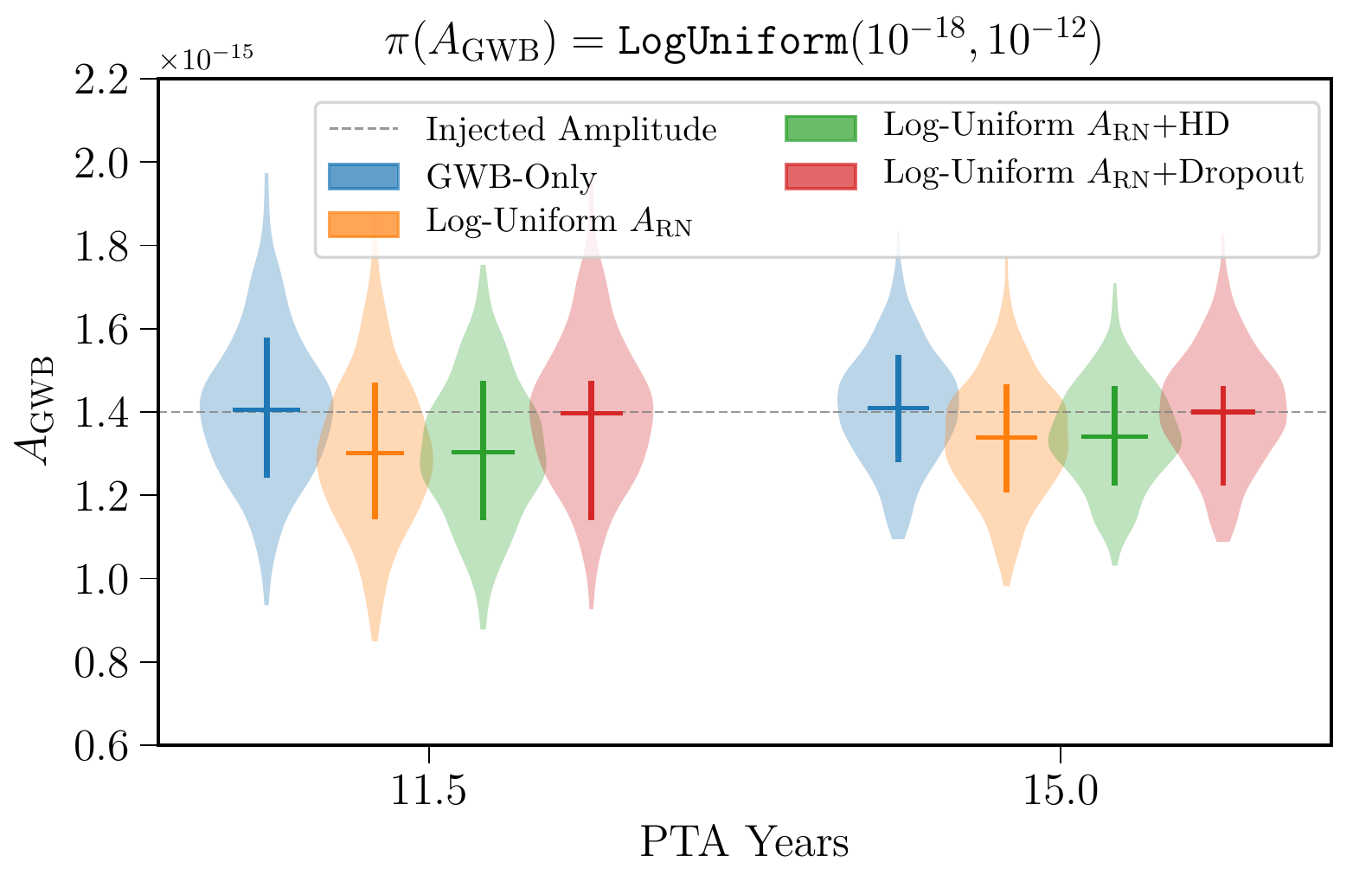}
\includegraphics[width=0.49\textwidth]{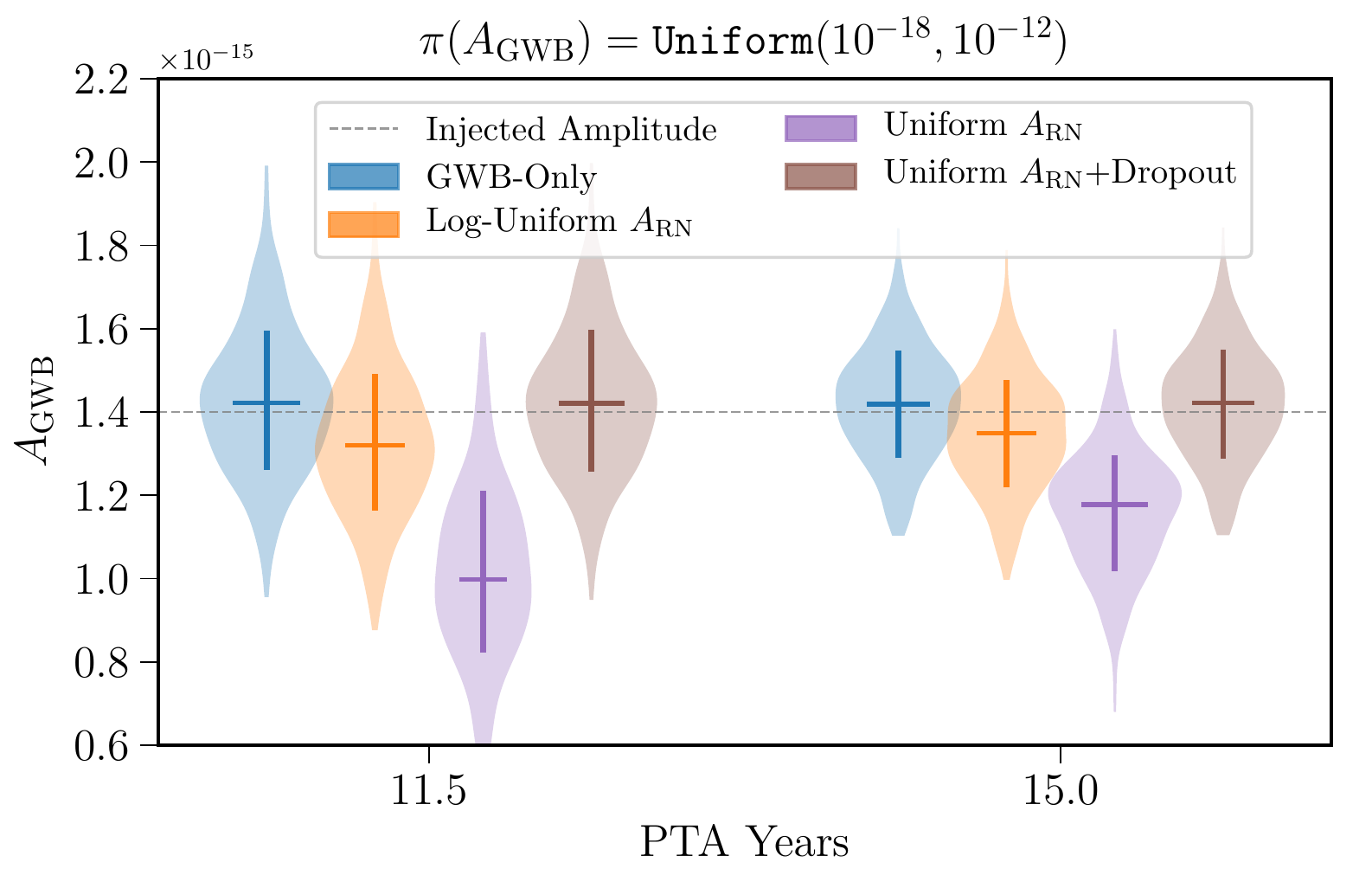}
\caption{Comparison of the distributions of median values of 
\Agwb for various signal+noise models and prior distributions, 
obtained from analyzing 400 realizations of the GWB+WN simulations.  
The left and right panels correspond to log-uniform and uniform
prior distributions for \Agwb.
The two groups of four violin plots per panel correspond to two 
different observation times (11.5 and 15~years).  
The horizontal bars show the median of the recovered median values of $A_{\rm GWB}$, 
while the vertical bars show the central 68\% credible interval
around that median value.  
The different color violin plots correspond to the different 
signal+noise models and prior distributions listed in Table~\ref{tab:models}.}
\label{fig:median_compare} 
\end{center}
\end{figure}

\begin{enumerate} 

\item 
As the signal to noise ratio increases the choice of prior on the signal amplitude
(in this case \Agwb) has little effect on the median value of \Agwb, which
is expected in a Bayesian analysis. 
This can be seen by comparing the GWB-only and 
\loguniform \Arn results (blue and orange violin plots, respectively) 
in the two panels of Figure~\ref{fig:median_compare}. 
The choice of prior on \Agwb does not
considerably change the distribution of median values.

\item 
The choice of prior on \Arn has a dramatic effect on the recovery of 
the median value of \Agwb. This can be seen by comparing the orange and
purple violin plots in the right-hand panel of \autoref{fig:median_compare}.
These correspond to log-uniform and uniform priors on $A_{\rm RN}$, 
respectively.

\item 
With log-uniform priors on both $A_{\rm GWB}$ and $A_{\rm RN}$, 
recovered median values of \Agwb are systematically lower
than the injected value $1.4\times 10^{-15}$ 
(orange violin plots in the left-hand panel of \autoref{fig:median_compare}).
This bias remains even if HD correlations are included in the signal
model (green violin plots in the left-hand panel of \autoref{fig:median_compare}).

\item The dropout model mitigates the effects of the \Arn prior on the posterior. 
It does that by including an intrinsic RN model only when it is really needed, returning results consistent with
the injected amplitude of the GWB
(red and brown violin plots, respectively, in the two panels of
\autoref{fig:median_compare}).
The difference between the brown and purple distributions shows that 
even \uniform
priors on {\it both} \Agwb and \Arn return fairly accurate median
values for \Agwb when the dropout method is used.  

\end{enumerate}

While the differences in the third point above seem fairly small
(i.e., the orange and green/ second and third violin plots in each set of the left-hand panel of
\autoref{fig:median_compare} are shifted lower by about $7\%$), 
these shifts can have fairly drastic results when considering the
interpretation of a single posterior from real data. Instead of
distributions of the median, one can tally the quantile position of
the injected value for each of the ``detection runs'' (which use
log-uniform priors for \Agwb). In the models where 
red noise is assumed for all pulsars, the injected value falls 
higher than a given credible interval 3 to 7 times more often than 
it falls lower than the same credible interval.
In other words, one is 3 to 7 times more likely to {\em underestimate} 
\Agwb than overestimate it. 
However, when the red noise dropout method is used, the discrepancy 
between the injected value falling higher or lower than the 
credible interval 
is reduced considerably, giving only a 20\% difference, corresponding 
to a factor of 1.2.

\begin{comment}
\begin{deluxetable*}{l|ccc|ccc}
\tablenum{1}
\tablecaption{Injected value of $A_{\rm GWB}$ and $\pi_{\rm
post}(A_{\rm GWB})$ credible intervals. Here we tabulate the location
of the injected value of $A_{\rm GWB}$ relative to the posteriors for
each realization. The columns show the percentage of times the
injected value for the GWB falls within or higher/lower than the
credible intervals calculated from the posterior probability densities
on $A_{\rm GWB}$.\label{tab:inj_val}}
\tablewidth{0pt}
\tablehead{
\nocolhead{Name} & \multicolumn{3}{c}{$68\%$ CI} & \multicolumn3c{$95\%$ CI} \\
\colhead{Model} & \colhead{\% within} & \colhead{\% lower} & \colhead{\% higher} &
\colhead{\% within} & \colhead{\% lower} & \colhead{\% higher}
}
%\decimals%colnumbers
\startdata
GWB_{\rm CRN} Only  & 61.5 & 20.75 & 17.75 & 89.0 & 5.5 & 5.5 \\
GWB_{\rm CRN} + RN & 57.75 & 9.25 & 33.0 & 89.0 & 1.25 & 9.75 \\
GWB_{\rm HD} + RN & 67.5 & 6.25 & 26.25 & 92.25 & 1.0 & 6.75 \\
GWB_{\rm CRN} + RN_{\rm DO} & 62.35 & 19.0  & 18.75 & 89.0 & 5.0 &  6.0 \\
GWB_{\rm CRN} + RN_{\rm DO} + INJ & 59.75 &  21.0 & 19.25 & 90.75 & 5.0 & 4.25 \\
\enddata
\tablecomments{GWB: GW background model	, RN: intrinsic red noise
model, HD: Hellings-Downs spatial correlations, DO: dropout analysis
on RN, INJ: RN injected into a handful of pulsars.}
\end{deluxetable*}
\end{comment}

%%%%%%%%%%%%%%%%%%%%%
\subsection{Effect of signal+noise models on GWB upper limits} 
\label{sec:ul_skew}

\begin{figure}[htb!]
\begin{center}
\includegraphics[width=0.49\textwidth]{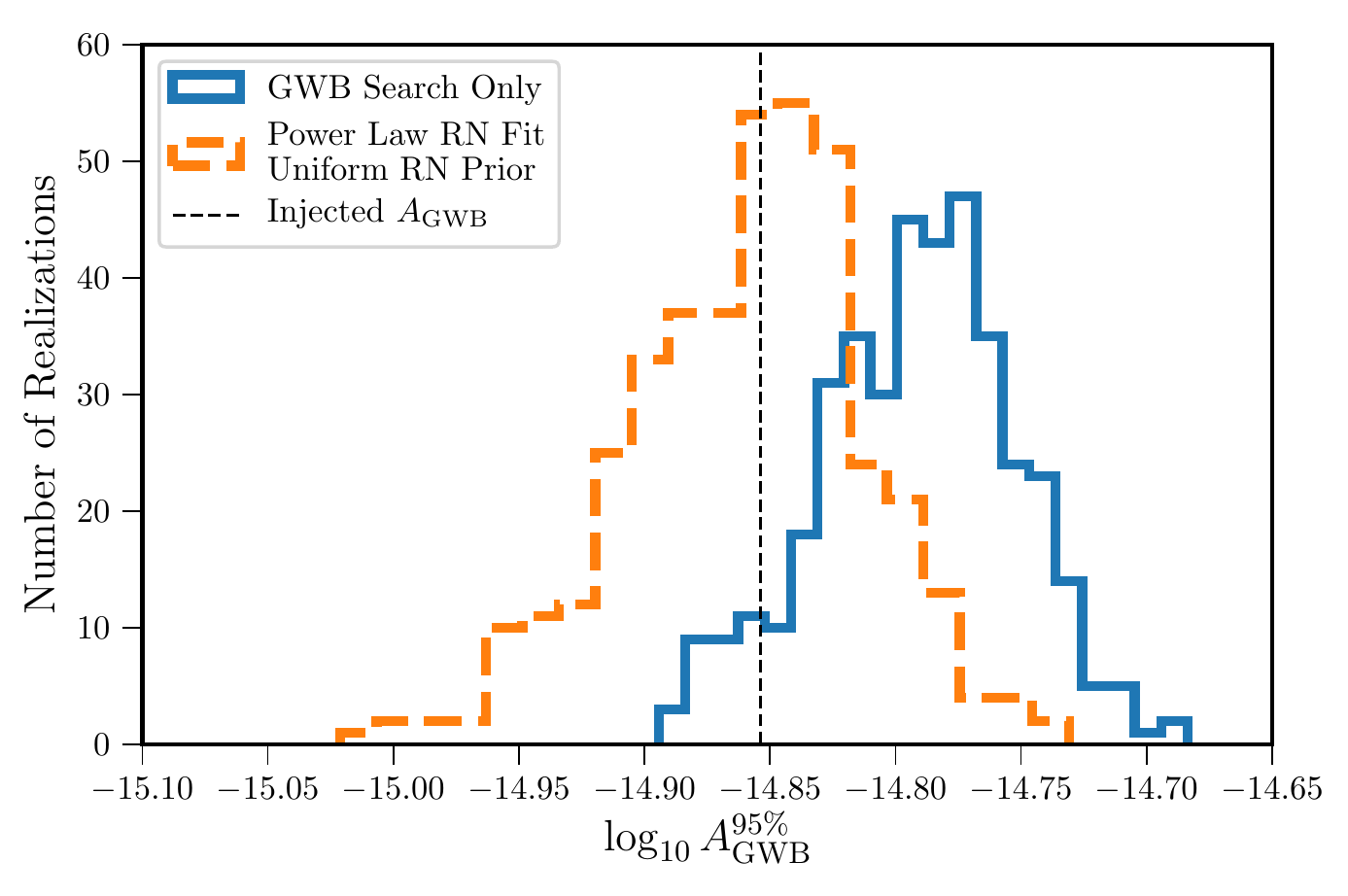}
\includegraphics[width=0.49\textwidth]{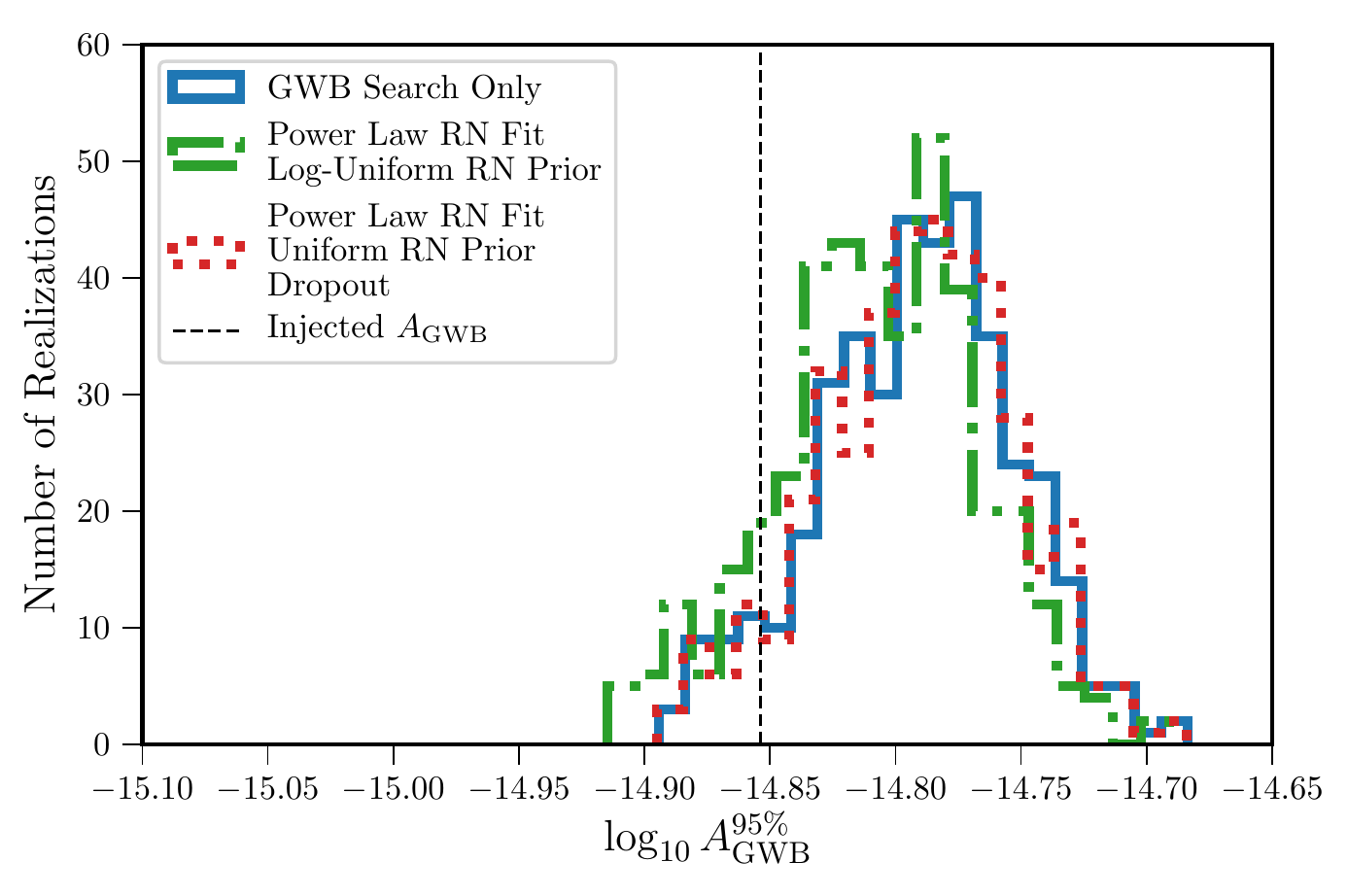}
\caption{Comparison of the distributions of the $95\%$ ULs for 
various signal+noise models and prior distributions, 
obtained from analyzing 400 realizations 
of the GWB+WN simulations. 
The left panel shows results for both the simple GWB-only 
signal+noise model (blue solid) and the standard PTA
UL analysis (orange dashed), 
which additionally includes red noise models for 
each pulsar with a \uniform\ prior on the amplitude \Arn. 
The right panel also shows results for a signal+noise model 
that uses log-uniform priors for \Arn (green dot-dashed), 
and a dropout model for uniform red noise priors (red dotted).
\label{fig:ul_compare2}} 
\end{center}
\end{figure}

We also determine the effects of the different signal+noise models and prior distributions on the 95\% UL, \Aul.
For all of the UL analyses, we use the standard convention of 
using a \uniform prior on \Agwb
\citep{Lentati:2016ygu,abb+16,abb+18b}.
For the GWB-only signal+noise model, we recover results for \Aul 
consistent with expectations (see the solid blue curve in either 
the left or right-hand panel of \autoref{fig:ul_compare2}).
As discussed in \autoref{sec:param_estim}, this is because the 
GWB-only signal+noise model agrees with that used to 
produce the simulated data.
The distribution of UL values calculated for the 400 realizations 
of the simulated data has values that are greater than the
injected value of the background roughly 93\% of the time 
(within error of the expected value of 95\%).
%  Error is sqrt(p*(1-p)/N)/P(x=x^p%), where the probability distribution is assumed Normal
%  Here with N=400, p=0.05 and P=0.10314, the estimated fractional error is 0.10565 or ~10.6%

We then perform analyses using the standard 
model for PTA GWB searches which includes an intrinsic red noise model for 
each pulsar. 
Recall that these models introduce $2N_{\rm pulsars}$ 
additional parameters (an amplitude, \Arn, and spectral index, 
$\gamma_{\rm RN}$, for each pulsar). 
To the best of our knowledge all
Bayesian PTA ULs to date have used uniform priors on $A_{\rm RN}$ and $A_{\rm GWB}$. 
From the dashed orange curve shown in the left-hand panel of 
\autoref{fig:ul_compare2}, we see that the distribution of 95\% ULs 
is shifted to significantly lower values, with basically {\em even
odds}, i.e., 50\%, that the calculated UL is above or below the injected value. 

It is worth pointing out that there is nothing wrong with the ULs produced by this procedure, as long as one is explicit about the model being used in the Bayesian analysis. However, if the signal+noise model that we are using differs greatly from what gave rise to the data, then statistical inferences can be systematically (and significantly) inaccurate.
The standard PTA UL analysis is based on the 
``conservative" assumption that all pulsars have
substantial red noise. 
In our simulations this assumption leads to individual pulsar red
noise models that are able to absorb a substantial amount of the
common red process, i.e., the GWB, and thus produce 
an overall {\em smaller} \Aul. 
The effects of the ``conservative'' assumption can be somewhat mitigated by using instead a
\loguniform prior on \Arn. 
This choice of prior decreases the bias, as one can 
see from the dot-dashed green distribution of \Aul in the 
right-hand plot of \autoref{fig:ul_compare2}, which has
87.25\% of its ULs above the injected value.
The \loguniform prior, however, is still part of a signal+noise model
that assumes that {\em all} pulsars have at least some level of measurable red noise.

As we have already seen in \autoref{sec:param_estim}, 
a better option is to use a red noise pulsar model in conjunction
with the dropout method, which 
allows the data to decide whether a given pulsar should be modeled
to include intrinsic red noise. 
The dropout analysis turns off the red noise models in almost all 
cases for this simple simulation of GWB+WN-only, returning us to 
results commensurate with the GWB-only search (see the dotted red 
distribution in the right-hand plot of \autoref{fig:ul_compare2},
which has 93.25\% of its ULs above the injected value).
The dropout method does this by using the threshold value to effectively set a prior on the 
presence (or absence) of intrinsic red noise in our pulsars, allowing the data to inform the choice of noise model. 

\begin{figure}[htb!]
\begin{center}
\includegraphics[width=0.8\textwidth]{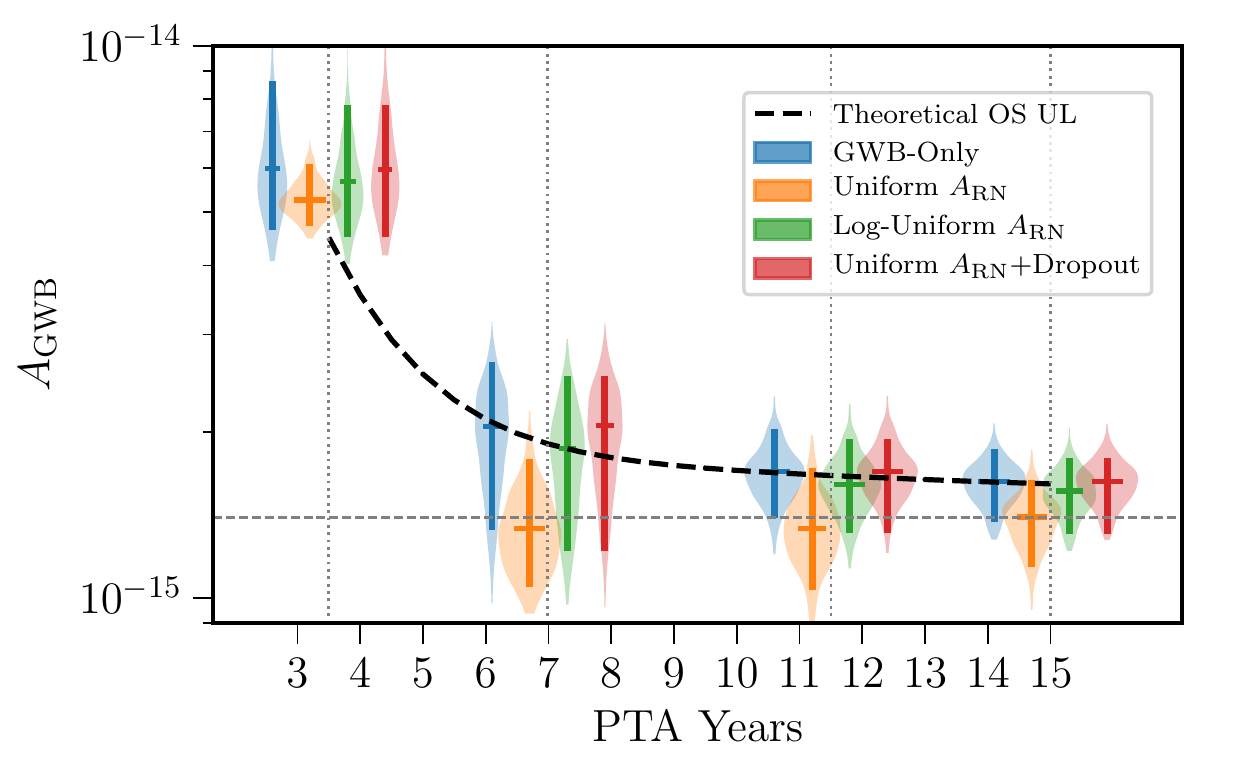}
\caption{Comparison of the distributions of the $95\%$ ULs for various
signal+noise models and prior distributions, 
obtained from analyzing 400 realizations of the GWB+WN simulations.  
The dashed black line shows the theoretical evolution of the UL based
on the frequentist optimal statistic, see~\autoref{eq:os_ul}.
The violin plots show the distributions of Bayesian 95\% ULs for four 
different signal+noise models and priors at four different time slices of the 
full dataset. The vertical bar within a violin plot shows
the central $90\%$ credible interval. The horizontal dashed gray line shows the injected amplitude of $1.4\times10^{-15}$.}
\label{fig:ul_compare}
\end{center}
\end{figure}
In \autoref{fig:ul_compare} we show the evolution of the UL as a
function of time for the various analyses. The injected \Agwb is
specifically chosen so that the GWB begins to have power greater than
the WN at the lowest frequencies when there is seven years of data.
Before this time the ULs are all still well above the injected value,
but as soon as one enters the intermediate signal
regime\footnote{Defined in \citet{sejr13} as beginning when the power
in the lowest frequency bin of the GWB is greater than that of the
WN.}, biases in the standard GWB UL analysis (orange violin plots) 
are clearly manifest.

In comparison, the distributions for the dropout analysis (red violin
plots) match those for the GWB-only search (blue violin plots), 
which also match the expected 95\%
confidence-level UL (black dashed line) 
predicted by the frequentist optimal
statistic (OS) developed in \citet{esvh13,sejr13,ccs+2015,vite18}. The
scaling laws of \citet{sejr13} are used to construct an analytic 
expression for \Aul as a function of time using \autoref{eq:os_ul}. 
Thus, just as we saw in \autoref{sec:param_estim}, the pulsar red noise
dropout analysis performs better than the standard PTA analysis
(uniform amplitude priors for both $A_{\rm GWB}$ and
$A_{\rm RN}$) for 95\% ULs as well.

%%%%%%%%%%%%%%%%%%%%%%%
\section{Application to more realistic data sets}
\label{sec:do_rn}

\begin{figure}[ht!]
\begin{center}
\includegraphics[width=\textwidth]{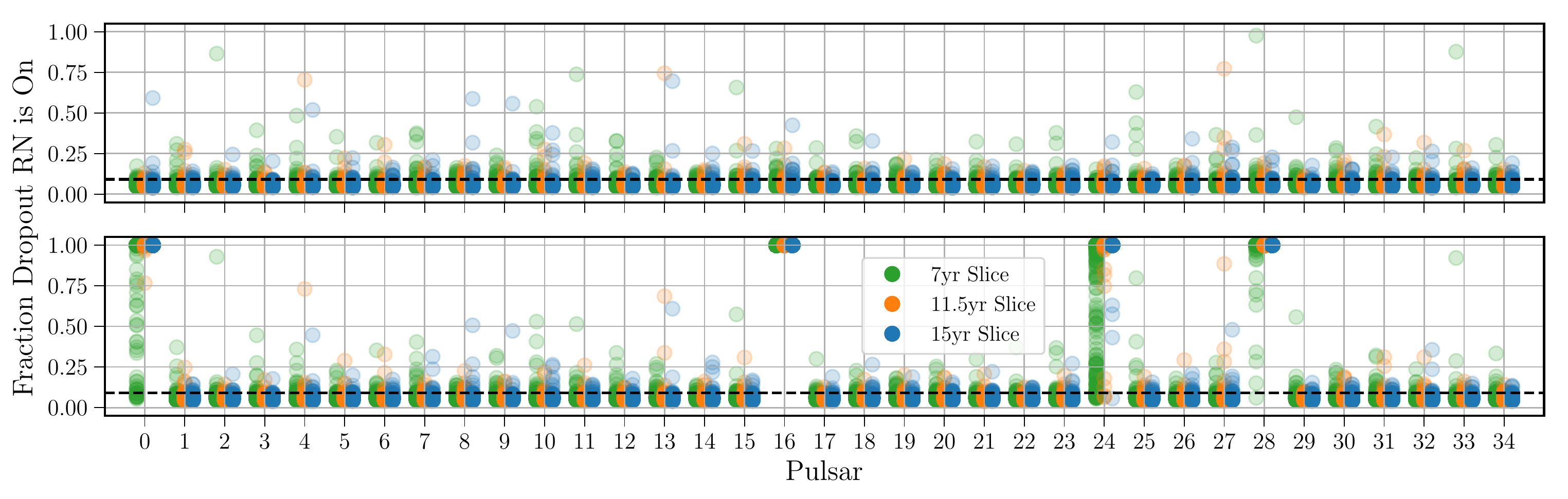}
\includegraphics[width=0.8\textwidth]{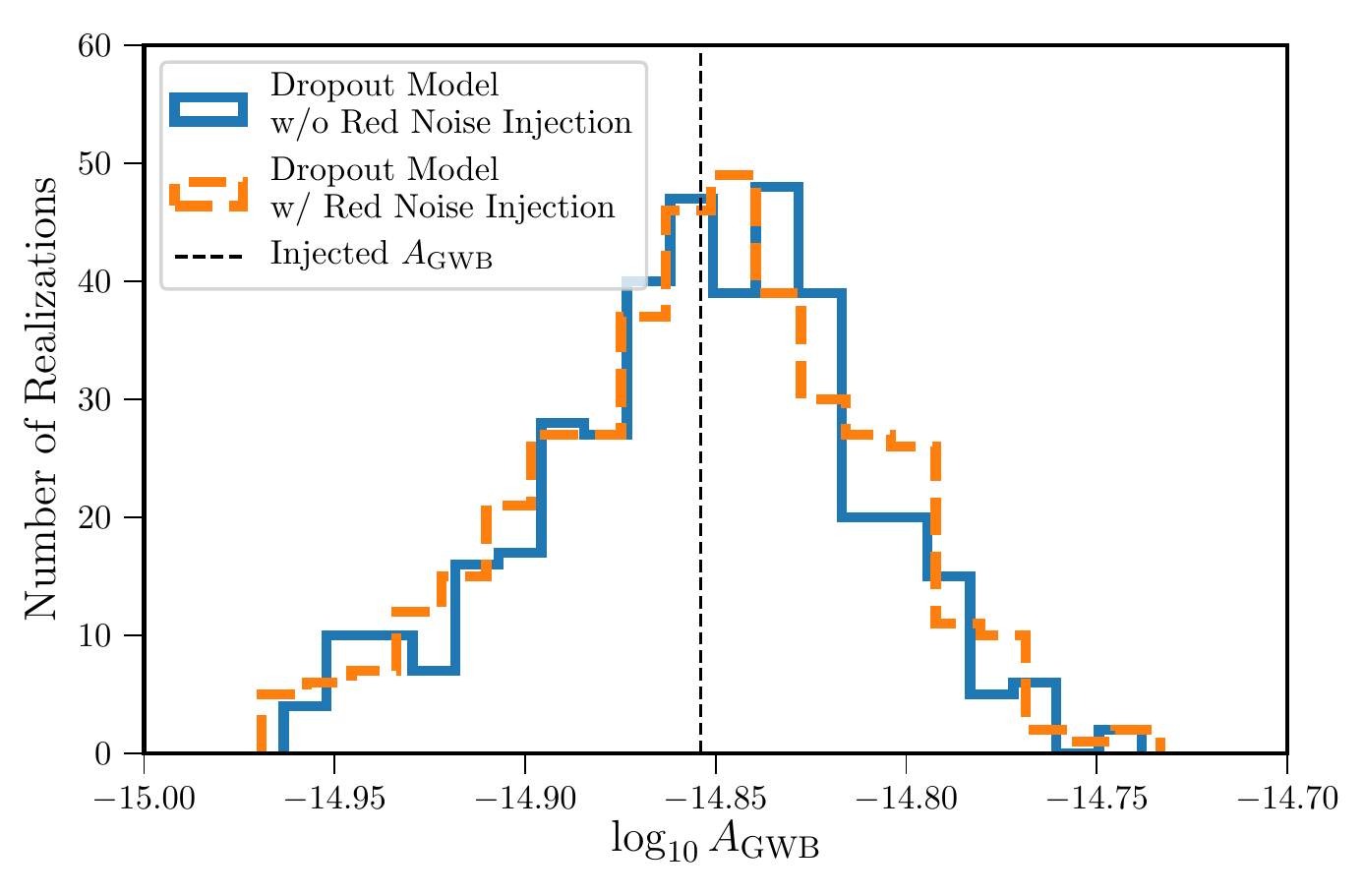}
\caption{Summary of dropout analyses for simulated dataset analyses.  
In the top two panels
each dot represents the fraction of the samples from a particular
realization/analysis when the red noise model is turned on;
dots below the dashed black line indicate that a particular pulsar's 
red noise model is disfavored.
The topmost panel shows the fraction turned on for the analysis of the 
GWB+WN only simulations.
The middle panel is for a simulation where 
additional red noise is injected into pulsars 0, 16, 24 and 28 with 
\Arn equal to ($10^{-15},3.4\times10^{-13},8\times10^{-15},7\times10^{-14}$)  and
spectral indices $\gamma_{\rm RN}$ equal to ($7,2,5,3$). 
The bottom panel shows the distribution of the median values for 
\Agwb for the dropout analysis
applied to both the GWB+WN-only and GWB+WN+RN simulations.} 
\label{fig:dropout1}
\end{center}
\end{figure}

As discussed earlier, the simple simulations studied in the 
previous sections only included WN and a red GWB signal. For these
simulations, the dropout model successfully turns off the 
red noise models in all pulsars, as can be seen in the top 
panel of \autoref{fig:dropout1}.
In this panel, the vertical height of the dots shows the 
fraction of time the red noise dropout model is turned on 
for a given realization at a given slice of the dataset. 
A dataset that is completely ambivalent about the presence 
of red noise would lie along the horizontal thin dashed line, i.e.,
the pulsars will have their red noise model turned on $1/11$th of the time, while the
presence of most dots below the line shows that red noise is
disfavored.

In order to assess the full abilities of the red noise dropout model,
another set of simulations were run where a handful of pulsars were
injected with red noise characteristic of that seen in real PTA
datasets. Various amplitudes and spectral indices of red noise were
injected and are detailed in the caption of \autoref{fig:dropout1}.
The middle panel shows the analogous results to the top panel, but
with the successful modeling of red noise in the pulsars where it has
been injected. Depending on the spectral index and amplitude, it may
take longer in the dataset to resolve the red noise in the pulsar, as
shown by the dependence of \autoref{eq:os_snr} on $\gamma$. However,
for the full 15 years of data, those pulsars with injected red noise
have red noise dropout models turned on through most steps of the
MCMC. The bottom panel of 
\autoref{fig:dropout1} shows the distribution of medians for \Agwb
from both dropout analyses. As one can see, the red noise dropout model
does just as well at estimating \Agwb whether there is red noise
present in some of the pulsars or not.

%%%%%%%%%%%%%%%%%%%%%%%%%%%%%%%%%%%%%%%%%%%%%%%%%%%%%%%
\section{Reassessing the NANOGrav 11-Year Analysis}
\label{sec:11yr}

\begin{figure}[ht!]
%\gridline{\fig{dropout_k_th1011_ng11yr_mod2a.pdf}{0.6\textwidth}{}
%\fig{ng11yr_A_gwb_dropout_rn_th1011.pdf}{0.4\textwidth}{}}
\begin{center}
\includegraphics[width=0.8\textwidth]{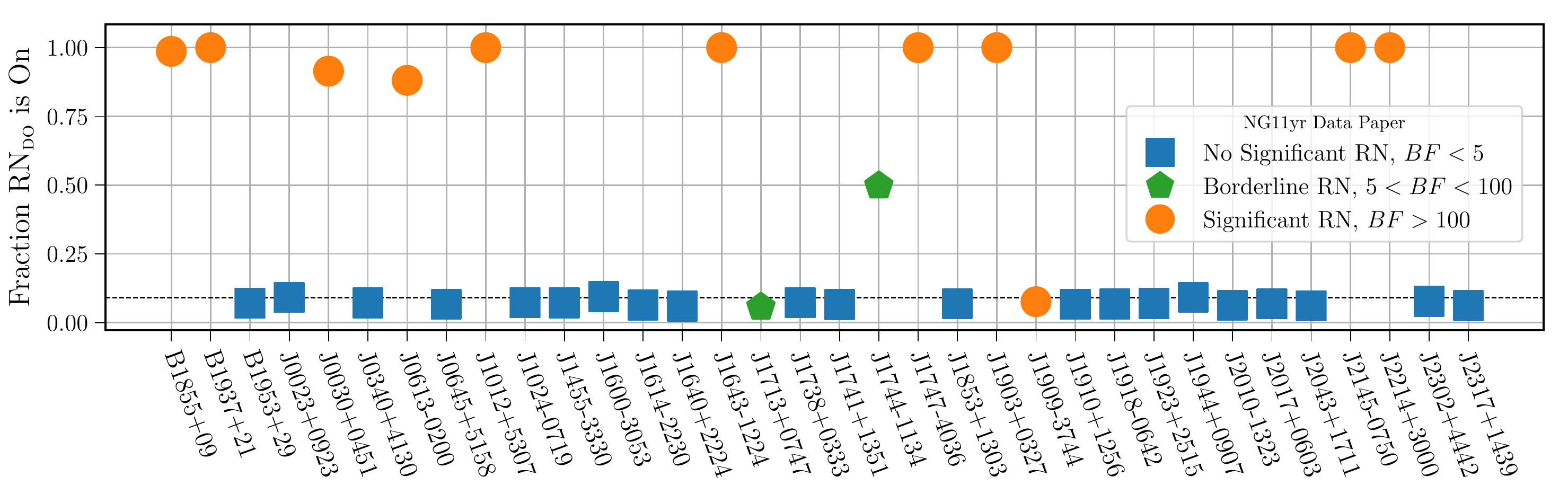}
\includegraphics[width=0.8\textwidth]{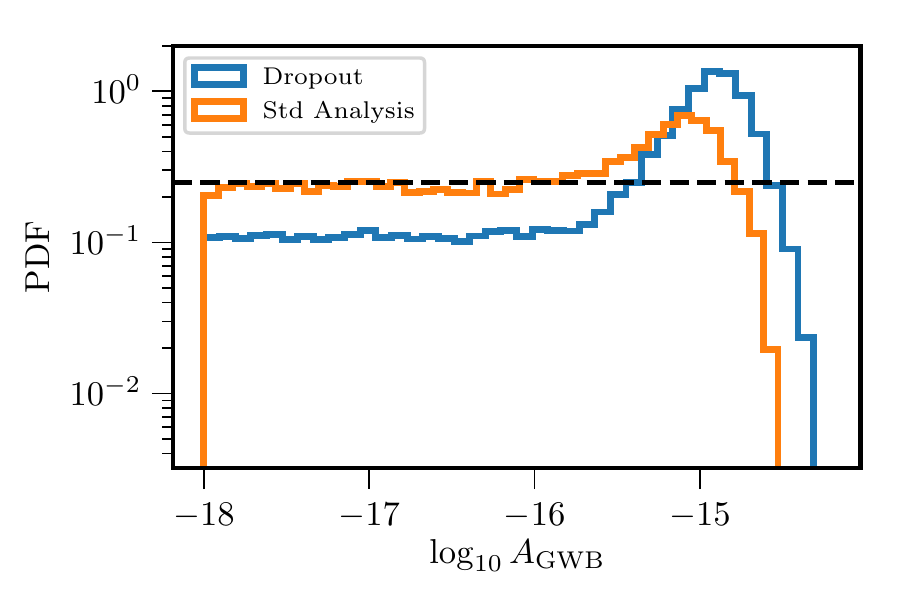}
\caption{Summary of the dropout analysis for the NANOGrav 11-year data.
In the top panel,
the various symbols show the fraction of samples where the red noise
model is turned on for a given pulsar, color coded by the red noise Bayes factor.  
The bottom panel compares the GWB
posteriors, including Hellings-Downs spatial correlations, for the NANOGrav
11-year data with (blue) and without
(orange) the red noise dropout analysis applied.
The maximum a~posteriori value of the GWB amplitude is 
$A_{\rm GWB}=1.41\times10^{-15}$ when the dropout analysis is
applied; it equals $1.1\times 10^{-15}$ without the dropout analysis.}
\label{fig:dropout_ng11yr}
\end{center}
\end{figure}

Finally we turn to real PTA data and apply the red noise dropout model
to the NANOGrav 11-year dataset \citep{abb+18a}.  In addition to
adding a dropout parameter for each of the pulsars, we also use the
{\sc BayesEphem} \citep{vts+20} solar system ephemeris model so that
our results are directly comparable to those of \citet{abb+18b}. This analysis includes
the Hellings-Downs spatial correlations for the GWB. 
Looking at the dropout parameters in the top panel of
\autoref{fig:dropout_ng11yr} we see that the analysis largely agrees
with the noise analysis used in \citet{abb+18a}. 
The only pulsar deemed significant in \citet{abb+18a}, where the red-noise model is turned off a majority
of the time during the dropout analysis is PSR J1909--3744.  Even though the red
noise model for this pulsar might return uninformative posteriors
for red noise when turned on in a normal analysis, it is still able to
absorb power that is better described by a common signal.

The posterior on \Agwb is compared to the standard PTA analysis
of the \citet{abb+18a} dataset in \autoref{fig:dropout_ng11yr}. Any
evidence for a detection using the dropout analysis, though slightly 
better, is still marginal.
However, as might be expected from the analyses in
\autoref{sec:param_estim}, the maximum a~posteriori (MAP) value for 
\Agwb using the dropout analysis is \Agwb$=1.4\times10^{-15}$, 
which is larger than that from the standard analysis, 
\Agwb$=1.1\times10^{-15}$, and more similar to the $95\%$ UL obtained in \citet{abb+18b}. We can re-weight the
samples in either of the analyses \citep{gelman2013bayesian} shown in
\autoref{fig:dropout_ng11yr} to obtain new $95\%$ ULs. Re-weighting the
samples from the standard PTA analysis is equivalent to the
\loguniform \Arn analysis discussed in \autoref{sec:ul_skew} and gives
\Aul$=2.1\times10^{-15}$. If instead we re-weight the samples from
the dropout analysis from \autoref{fig:dropout_ng11yr}, we obtain
results comparable to the dropout analysis from \autoref{sec:ul_skew},
which appears to be the most trustworthy model examined here for
obtaining ULs. This gives an UL for the NANOGrav 11-year dataset of
\Aul$=3.0\times10^{-15}$, {\em which is more than twice as large} 
as the UL quoted in \citet{abb+18b}. This result is dependent on the threshold set for the dropout parameter, as discussed in \autoref{sec:dropout}, and should not be taken as a concrete astrophysical result, but rather an example of how the GWB statistics can shift when using this new model. For the most up-to-date GWB results see Arzoumanian, et al., Submitted. 

\section{Conclusions}
\label{sec:conclusions}
Here we have shown explicitly how the choice of
prior on \Arn, and indeed whether pulsars have red noise models at
all, can have unanticipated consequences on the statistics of \Agwb.
In a simulated data set with WN and a GWB, the effect of a standard GWB search with uniform priors for the pulsar intrinsic red noise amplitudes on \Aul is drastic, returning a $95\%$ UL {\it lower} than
the injected value in about half of all realizations. 
%A uniform prior on the amplitude is biased high, in general, which informs the idea that the uniform prior on \Agwb is more``conservative''. 
As we have seen, putting a
uniform prior on \Arn biases the noise model to steal power from the GWB model. 
%This not only undoes,
%but overwhelms the "bias high" on \Agwb. 
The biases of other estimators, such as the conditional median,
that occur for parameter estimation are smaller, but still show a
consistent shift in the same direction, i.e., to smaller values of \Agwb. 
We have implemented a simple solution to these problems---the so-called dropout method---which is a flexible red noise  model for pulsars that allows the intrinsic red noise model in each pulsar to be turned off during the course of the Bayesian analysis if there is not sufficient evidence in the data to warrant its presence. 

In light of the offsets in parameter estimation uncovered by this work, it is worth revisiting constraints inferred on the SMBBH population from PTA datasets.
The impact of the choice of prior on the intrinsic red noise
amplitudes can be seen directly by comparing the astrophysical
interpretation done in NANOGrav's 9-year GWB constraint paper \citep{abb+16}, which
used uniform priors on red noise amplitude, with that done in its
11-year GWB constraint paper \citep{abb+18b}, which used log-uniform priors. Even
though the 11-year constraint on \Agwb is a smaller value, the
astrophysical inference is less constraining. This is partially due to
the differing models and analysis techniques. However, viewed through the lens of this work, the weakening of constraints, specifically on the $M-M_\mathrm{bulge}$ relationship, were certainly impacted by the choice of prior used for the red noise amplitudes.
Beyond direct constraints using \Agwb, previously reported \Aul upper limits have been used in concert with electromagnetic observations to make statements about various SMBBH population models \citep{hss+18,shk+18}.
Moving forward, astrophysical statements derived from past PTA constraints on \Agwb will need to be more cautious in assessing the strength of their inference.

The idea that credible intervals and parameter estimation are
dependent on the choice of model and priors is a common refrain in
Bayesian statistics. As a result data analysts should strive to produce models and priors that robustly represent and quantify the underlying physical processes being investigated. 

\acknowledgments

This work was supported by the NSF NANOGrav Physics Frontier Center (NSF PHY-1430284). The authors thank Paul Baker, Michael Lam, Timothy Pennucci, Stephen Taylor, Michele Vallisneri and Sarah Vigeland for useful discussions and detailed comments on early drafts of this manuscript.  Part of this research was carried out at the Jet Propulsion Laboratory, California Institute of Technology, under a contract with the National Aeronautics and Space Administration. Joseph Simon acknowledges support from the JPL RTD program. Lastly, we are grateful for computational resources provided by the Leonard E Parker Center for Gravitation, Cosmology and Astrophysics at the University of Wisconsin-Milwaukee (NSF PHY-1626190).

\vspace{5mm}
\facilities{Arecibo, GBT}

\software{astropy \citep{astropy},
	      libstempo \citep{libstempo},
	      Tempo2 \citep{tempo2},
          la\_forge, 
          {\sc Enterprise} \citep{enterprise},
          enterprise\_extensions,
          PTMCMCSampler \citep{ptmcmc}
          }

\appendix

\section{Time evolution of a frequentist GWB Upper Limit} 
\label{sec:os}

Here we derive an expression for the expected value of the
frequentist 95\% confidence-level UL calculated from the 
optimal statistic~\citep{esvh13,sejr13,ccs+2015,vite18}.
Although this is a frequentist UL, it provides a good 
analytic approximation to the Bayesian 95\% ULs calculated in this paper.

The expected signal-to-noise ratio $\rho$ derived from the optimal 
statistic can be written in the
frequency domain as \citep{ccs+2015}: 
\begin{equation}
\rho\equiv\sqrt{\left<\rho^2\right>}= \left(2T\sum_{IJ} \chi^2_{IJ}
\int^{f_H}_{f_L}  df
\frac{{P}_g^2(f)}{{P}_I(f){P}_J(f)}\right)^{1/2}\,,
\end{equation} 
where the indices $I$ and $J$ label the individual pulsars, 
${P}_I(f)$ is the total auto-correlated power spectral density for pulsar $I$, 
${P}_g(f)$ is the power spectral density for the GWB, 
$T$ is the time span of the data, and $\chi_{IJ}$
are the overlap reduction function coefficients, here assumed to be
the quadrupolar spatial correlations induced by a GWB \citep{hd83}. 

The expression for the signal-to-noise ratio can be simplified
considerably for the main set of simulations considered in this work
where all of the pulsars have the same level of white noise,
cadence and observing time span, and there is no red noise injected
into the pulsars:
\begin{equation}\label{eq:os_snr}
\rho = \left( 2T\sum_{IJ} \chi^2_{IJ} 
\int^{f_H}_{f_L}  df
\frac{b^2f^{-2\gamma}}{\left(bf^{-\gamma}+2\sigma^2 \Delta
t\right)^{2}}\right)^{1/2}.
\end{equation}
Here $\sigma$ is the TOA error value for all the pulsars, $\Delta
t$ is the sampling period (the cadence), and $b$ subsumes various constants\footnote{Notice the difference in minus sign of the exponent
corrected from \citet{hazboun:2020slice}.},
\begin{equation}
b\equiv \frac{A_{\rm GWB}^2}{12\pi^2}\left(\frac{1}{f_{\rm yr}}\right)^{-\gamma+3}\,.
\end{equation}
One can relate the aforementioned scaling laws to an UL by using the
complimentary error function
\citep{Allen-Romano:1999,hazboun:2020slice}:
\begin{equation}
A^2_{\rm UL}=\hat{A}^2_{\rm GWB}+\frac{\sqrt{2}\sigma_0}{\sqrt{T}} 
\;{\rm erfc}^{-1}\left[2\left(1-\mu\right)\right]\,,
\qquad
\rho \equiv \frac{A^2_{\rm GWB}}{\sigma_0/\sqrt{T}}\,,
\label{e:AUL}
\end{equation}
where $\mu$ is the confidence level
(e.g., $\mu = 0.95$ for a $95\%$ confidence-level UL), 
and $\sigma_0$ is the effective noise level defined in terms
of $\rho$, $A_{\rm GWB}$, and $T$.
The expectation value of \autoref{e:AUL} yields
\begin{equation}\label{eq:os_ul}
\left<A^2_{\rm UL}\right>
=A^2_{\rm GWB}\left(1+\frac{\sqrt{2}\; {\rm
erfc}^{-1}\left[2\left(1-\mu\right)\right]}{\rho}\right)\,.
\end{equation}
For the simple GWB+WN simulations studied in the majority of 
this paper, this UL is calculated analytically and plotted in
\autoref{fig:ul_compare} as the dashed black line.

\bibliography{sims_bib}{}

\begin{thebibliography}{}
\expandafter\ifx\csname natexlab\endcsname\relax\def\natexlab#1{#1}\fi
\providecommand{\url}[1]{\href{#1}{#1}}
\providecommand{\dodoi}[1]{doi:~\href{http://doi.org/#1}{\nolinkurl{#1}}}
\providecommand{\doeprint}[1]{\href{http://ascl.net/#1}{\nolinkurl{http://ascl.net/#1}}}
\providecommand{\doarXiv}[1]{\href{https://arxiv.org/abs/#1}{\nolinkurl{https://arxiv.org/abs/#1}}}

\bibitem[{{Aggarwal} {et~al.}(2019){Aggarwal}, {Arzoumanian}, {Baker},
  {Brazier}, {Brinson}, {Brook}, {Burke-Spolaor}, {Chatterjee}, {Cordes},
  {Cornish}, {Crawford}, {Crowter}, {Cromartie}, {DeCesar}, {Demorest},
  {Dolch}, {Ellis}, {Ferdman}, {Ferrara}, {Fonseca}, {Garver-Daniels},
  {Gentile}, {Hazboun}, {Holgado}, {Huerta}, {Islo}, {Jennings}, {Jones},
  {Jones}, {Kaiser}, {Kaplan}, {Kelley}, {Key}, {Lam}, {Lazio}, {Levin},
  {Lorimer}, {Luo}, {Lynch}, {Madison}, {McLaughlin}, {McWilliams},
  {Mingarelli}, {Ng}, {Nice}, {Pennucci}, {Pol}, {Ransom}, {Ray}, {Siemens},
  {Simon}, {Spiewak}, {Stairs}, {Stinebring}, {Stovall}, {Swiggum}, {Taylor},
  {Turner}, {Vallisneri}, {van Haasteren}, {Vigeland }, {Witt}, {Zhu}, \&
  {NANOGrav Collaboration}}]{aab+19}
{Aggarwal}, K., {Arzoumanian}, Z., {Baker}, P.~T., {et~al.} 2019, \apj, 880,
  116, \dodoi{10.3847/1538-4357/ab2236}

\bibitem[{{Aggarwal} {et~al.}(2020){Aggarwal}, {Arzoumanian}, {Baker},
  {Brazier}, {Brook}, {Burke-Spolaor}, {Chatterjee}, {Cordes}, {Cornish},
  {Crawford}, {Cromartie}, {Crowter}, {DeCesar}, {Demorest}, {Dolch}, {Ellis},
  {Ferdman}, {Ferrara}, {Fonseca}, {Garver-Daniels}, {Gentile}, {Good},
  {Hazboun}, {Holgado}, {Huerta}, {Islo}, {Jennings}, {Jones}, {Jones},
  {Kaplan}, {Kelley}, {Key}, {Lam}, {Lazio}, {Levin}, {Lorimer}, {Luo},
  {Lynch}, {Madison}, {McLaughlin}, {McWilliams}, {Mingarelli}, {Ng}, {Nice},
  {Pennucci}, {Pol}, {Ransom}, {Ray}, {Siemens}, {Simon}, {Spiewak}, {Stairs},
  {Stinebring}, {Stovall}, {Swiggum}, {Taylor}, {Vallisneri}, {van Haasteren},
  {Vigeland}, {Witt}, \& {Zhu}}]{aab+20}
---. 2020, \apj, 889, 38, \dodoi{10.3847/1538-4357/ab6083}

\bibitem[{{Allen} \& {Romano}(1999)}]{Allen-Romano:1999}
{Allen}, B., \& {Romano}, J.~D. 1999, Physical Review D, 59, 102001,
  \dodoi{10.1103/PhysRevD.59.102001}

\bibitem[{{Arzoumanian} {et~al.}(2014){Arzoumanian}, {Brazier},
  {Burke-Spolaor}, {Chamberlin}, {Chatterjee}, {Cordes}, {Demorest}, {Deng},
  {Dolch}, {Ellis}, {Ferdman}, {Garver-Daniels}, {Jenet}, {Jones}, {Kaspi},
  {Koop}, {Lam}, {Lazio}, {Lommen}, {Lorimer}, {Luo}, {Lynch}, {Madison},
  {McLaughlin}, {McWilliams}, {Nice}, {Palliyaguru}, {Pennucci}, {Ransom},
  {Sesana}, {Siemens}, {Stairs}, {Stinebring}, {Stovall}, {Swiggum},
  {Vallisneri}, {van Haasteren}, {Wang}, {Zhu}, \& {NANOGrav
  Collaboration}}]{abb+14}
{Arzoumanian}, Z., {Brazier}, A., {Burke-Spolaor}, S., {et~al.} 2014, \apj,
  794, 141, \dodoi{10.1088/0004-637X/794/2/141}

\bibitem[{Arzoumanian {et~al.}(2016)Arzoumanian, Brazier, Burke-Spolaor,
  Chamberlin, Chatterjee, Christy, Cordes, Cornish, Crowter, Demorest, Deng,
  Dolch, Ellis, Ferdman, Fonseca, Garver-Daniels, Gonzalez, Jenet, Jones,
  Jones, Kaspi, Koop, Lam, Lazio, Levin, Lommen, Lorimer, Luo, Lynch, Madison,
  Mclaughlin, Mcwilliams, Mingarelli, Nice, Palliyaguru, Pennucci, Ransom,
  Sampson, Sanidas, Sesana, Siemens, Simon, Stairs, Stinebring, Stovall,
  Swiggum, Taylor, Vallisneri, van Haasteren, Wang, Zhu, \&
  Collaboration}]{abb+16}
Arzoumanian, Z., Brazier, A., Burke-Spolaor, S., {et~al.} 2016, \apj, 821, 13

\bibitem[{{Arzoumanian} {et~al.}(2018{\natexlab{a}}){Arzoumanian}, {Baker},
  {Brazier}, {Burke-Spolaor}, {Chamberlin}, {Chatterjee}, {Christy}, {Cordes},
  {Cornish}, {Crawford}, {Thankful Cromartie}, {Crowter}, {DeCesar},
  {Demorest}, {Dolch}, {Ellis}, {Ferdman}, {Ferrara}, {Folkner}, {Fonseca},
  {Garver-Daniels}, {Gentile}, {Haas}, {Hazboun}, {Huerta}, {Islo}, {Jones},
  {Jones}, {Kaplan}, {Kaspi}, {Lam}, {Lazio}, {Levin}, {Lommen}, {Lorimer},
  {Luo}, {Lynch}, {Madison}, {McLaughlin}, {McWilliams}, {Mingarelli}, {Ng},
  {Nice}, {Park}, {Pennucci}, {Pol}, {Ransom}, {Ray}, {Rasskazov}, {Siemens},
  {Simon}, {Spiewak}, {Stairs}, {Stinebring}, {Stovall}, {Swiggum}, {Taylor},
  {Vallisneri}, {van Haasteren}, {Vigeland}, {Zhu}, \& {NANOGrav
  Collaboration}}]{abb+18b}
{Arzoumanian}, Z., {Baker}, P.~T., {Brazier}, A., {et~al.} 2018{\natexlab{a}},
  \apj, 859, 47, \dodoi{10.3847/1538-4357/aabd3b}

\bibitem[{{Arzoumanian} {et~al.}(2018{\natexlab{b}}){Arzoumanian}, {Brazier},
  {Burke-Spolaor}, {Chamberlin}, {Chatterjee}, {Christy}, {Cordes}, {Cornish},
  {Crawford}, {Thankful Cromartie}, {Crowter}, {DeCesar}, {Demorest}, {Dolch},
  {Ellis}, {Ferdman}, {Ferrara}, {Fonseca}, {Garver-Daniels}, {Gentile},
  {Halmrast}, {Huerta}, {Jenet}, {Jessup}, {Jones}, {Jones}, {Kaplan}, {Lam},
  {Lazio}, {Levin}, {Lommen}, {Lorimer}, {Luo}, {Lynch}, {Madison}, {Matthews},
  {McLaughlin}, {McWilliams}, {Mingarelli}, {Ng}, {Nice}, {Pennucci}, {Ransom},
  {Ray}, {Siemens}, {Simon}, {Spiewak}, {Stairs}, {Stinebring}, {Stovall},
  {Swiggum}, {Taylor}, {Vallisneri}, {van Haasteren}, {Vigeland}, {Zhu}, \&
  {NANOGrav Collaboration}}]{abb+18a}
{Arzoumanian}, Z., {Brazier}, A., {Burke-Spolaor}, S., {et~al.}
  2018{\natexlab{b}}, \apjs, 235, 37, \dodoi{10.3847/1538-4365/aab5b0}

\bibitem[{{Astropy Collaboration} {et~al.}(2018){Astropy Collaboration},
  {Price-Whelan}, {Sip{\H o}cz}, {G{\"u}nther}, {Lim}, {Crawford}, {Conseil},
  {Shupe}, {Craig}, {Dencheva}, {Ginsburg}, {VanderPlas}, {Bradley},
  {P{\'e}rez-Su{\'a}rez}, {de Val-Borro}, {Aldcroft}, {Cruz}, {Robitaille},
  {Tollerud}, {Ardelean}, {Babej}, {Bach}, {Bachetti}, {Bakanov}, {Bamford},
  {Barentsen}, {Barmby}, {Baumbach}, {Berry}, {Biscani}, {Boquien}, {Bostroem},
  {Bouma}, {Brammer}, {Bray}, {Breytenbach}, {Buddelmeijer}, {Burke},
  {Calderone}, {Cano Rodr{\'{\i}}guez}, {Cara}, {Cardoso}, {Cheedella},
  {Copin}, {Corrales}, {Crichton}, {D'Avella}, {Deil}, {Depagne}, {Dietrich},
  {Donath}, {Droettboom}, {Earl}, {Erben}, {Fabbro}, {Ferreira}, {Finethy},
  {Fox}, {Garrison}, {Gibbons}, {Goldstein}, {Gommers}, {Greco}, {Greenfield},
  {Groener}, {Grollier}, {Hagen}, {Hirst}, {Homeier}, {Horton}, {Hosseinzadeh},
  {Hu}, {Hunkeler}, {Ivezi{\'c}}, {Jain}, {Jenness}, {Kanarek}, {Kendrew},
  {Kern}, {Kerzendorf}, {Khvalko}, {King}, {Kirkby}, {Kulkarni}, {Kumar},
  {Lee}, {Lenz}, {Littlefair}, {Ma}, {Macleod}, {Mastropietro}, {McCully},
  {Montagnac}, {Morris}, {Mueller}, {Mumford}, {Muna}, {Murphy}, {Nelson},
  {Nguyen}, {Ninan}, {N{\"o}the}, {Ogaz}, {Oh}, {Parejko}, {Parley}, {Pascual},
  {Patil}, {Patil}, {Plunkett}, {Prochaska}, {Rastogi}, {Reddy Janga},
  {Sabater}, {Sakurikar}, {Seifert}, {Sherbert}, {Sherwood-Taylor}, {Shih},
  {Sick}, {Silbiger}, {Singanamalla}, {Singer}, {Sladen}, {Sooley},
  {Sornarajah}, {Streicher}, {Teuben}, {Thomas}, {Tremblay}, {Turner},
  {Terr{\'o}n}, {van Kerkwijk}, {de la Vega}, {Watkins}, {Weaver}, {Whitmore},
  {Woillez}, {Zabalza}, \& {Astropy Contributors}}]{astropy}
{Astropy Collaboration}, {Price-Whelan}, A.~M., {Sip{\H o}cz}, B.~M., {et~al.}
  2018, \aj, 156, 123, \dodoi{10.3847/1538-3881/aabc4f}

\bibitem[{{Blandford} {et~al.}(1984){Blandford}, {Narayan}, \&
  {Romani}}]{blandford+1984}
{Blandford}, R., {Narayan}, R., \& {Romani}, R.~W. 1984, Journal of
  Astrophysics and Astronomy, 5, 369, \dodoi{10.1007/BF02714466}

\bibitem[{{Burke-Spolaor} {et~al.}(2018){Burke-Spolaor}, {Taylor}, {Charisi},
  {Dolch}, {Hazboun}, {Holgado}, {Kelley}, {Lazio}, {Madison}, {McMann},
  {Mingarelli}, {Rasskazov}, {Siemens}, {Simon}, \& {Smith}}]{spolaor:2018}
{Burke-Spolaor}, S., {Taylor}, S.~R., {Charisi}, M., {et~al.} 2018, arXiv
  e-prints, arXiv:1811.08826.
\newblock \doarXiv{1811.08826}

\bibitem[{Callister {et~al.}(2017)Callister, Biscoveanu, Christensen, Isi,
  Matas, Minazzoli, Regimbau, Sakellariadou, Tasson, \&
  Thrane}]{Callister:2017}
Callister, T., Biscoveanu, A.~S., Christensen, N., {et~al.} 2017, Phys. Rev. X,
  7, 041058, \dodoi{10.1103/PhysRevX.7.041058}

\bibitem[{Carlin \& Chib(1995)}]{cc95}
Carlin, B.~P., \& Chib, S. 1995, Journal of the Royal Statistical Society.
  Series B (Methodological), 57, 473.
\newblock \url{http://www.jstor.org/stable/2346151}

\bibitem[{{Chamberlin} {et~al.}(2015){Chamberlin}, {Creighton}, {Siemens},
  {Demorest}, {Ellis}, {Price}, \& {Romano}}]{ccs+2015}
{Chamberlin}, S.~J., {Creighton}, J.~D.~E., {Siemens}, X., {et~al.} 2015, \prd,
  91, 044048, \dodoi{10.1103/PhysRevD.91.044048}

\bibitem[{{Champion} {et~al.}(2010){Champion}, {Hobbs}, {Manchester},
  {Edwards}, {Backer}, {Bailes}, {Bhat}, {Burke-Spolaor}, {Coles}, {Demorest},
  {Ferdman}, {Folkner}, {Hotan}, {Kramer}, {Lommen}, {Nice}, {Purver},
  {Sarkissian}, {Stairs}, {van Straten}, {Verbiest}, \&
  {Yardley}}]{champion2010}
{Champion}, D.~J., {Hobbs}, G.~B., {Manchester}, R.~N., {et~al.} 2010, \apjl,
  720, L201, \dodoi{10.1088/2041-8205/720/2/L201}

\bibitem[{Cordes(2013)}]{cordes2013}
Cordes, J.~M. 2013, Classical and Quantum Gravity, 30, 224002.
\newblock \url{http://stacks.iop.org/0264-9381/30/i=22/a=224002}

\bibitem[{{Cordes} \& {Downs}(1985)}]{cordes1985}
{Cordes}, J.~M., \& {Downs}, G.~S. 1985, \apjs, 59, 343, \dodoi{10.1086/191076}

\bibitem[{{Cordes} \& {Shannon}(2010)}]{cordes2010}
{Cordes}, J.~M., \& {Shannon}, R.~M. 2010, arXiv e-prints, arXiv:1010.3785.
\newblock \doarXiv{1010.3785}

\bibitem[{{Demorest} {et~al.}(2013){Demorest}, {Ferdman}, {Gonzalez}, {Nice},
  {Ransom}, {Stairs}, {Arzoumanian}, {Brazier}, {Burke-Spolaor}, {Chamberlin},
  {Cordes}, {Ellis}, {Finn}, {Freire}, {Giampanis}, {Jenet}, {Kaspi}, {Lazio},
  {Lommen}, {McLaughlin}, {Palliyaguru}, {Perrodin}, {Shannon}, {Siemens},
  {Stinebring}, {Swiggum}, \& {Zhu}}]{dfg+13}
{Demorest}, P.~B., {Ferdman}, R.~D., {Gonzalez}, M.~E., {et~al.} 2013, 762, 94,
  \dodoi{10.1088/0004-637X/762/2/94}

\bibitem[{{Detweiler}(1979)}]{det79}
{Detweiler}, S. 1979, \apj, 234, 1100, \dodoi{10.1086/157593}

\bibitem[{Dickey(1971)}]{dickey1971}
Dickey, J.~M. 1971, The Annals of Mathematical Statistics, 42, 204.
\newblock \url{http://www.jstor.org/stable/2958475}

\bibitem[{Efron(2015)}]{efron:2015}
Efron, B. 2015, Journal of the Royal Statistical Society. Series B (Statistical
  Methodology), 77, 617.
\newblock \url{http://www.jstor.org/stable/24774821}

\bibitem[{Ellis \& van Haasteren(2017)}]{ptmcmc}
Ellis, J., \& van Haasteren, R. 2017, jellis18/PTMCMCSampler: Official Release,
  \dodoi{10.5281/zenodo.1037579}

\bibitem[{{Ellis} \& {Cornish}(2016)}]{ellis:2016}
{Ellis}, J.~A., \& {Cornish}, N.~J. 2016, \prd, 93, 084048,
  \dodoi{10.1103/PhysRevD.93.084048}

\bibitem[{{Ellis} {et~al.}(2013){Ellis}, {Siemens}, \& {van
  Haasteren}}]{esvh13}
{Ellis}, J.~A., {Siemens}, X., \& {van Haasteren}, R. 2013, 769, 63,
  \dodoi{10.1088/0004-637X/769/1/63}

\bibitem[{{Ellis} {et~al.}(2019){Ellis}, {Vallisneri}, {Taylor}, \&
  {Baker}}]{enterprise}
{Ellis}, J.~A., {Vallisneri}, M., {Taylor}, S.~R., \& {Baker}, P.~T. 2019,
  {ENTERPRISE: Enhanced Numerical Toolbox Enabling a Robust PulsaR Inference
  SuitE}.
\newblock \doeprint{1912.015}

\bibitem[{{Foster} \& {Backer}(1990)}]{fb90}
{Foster}, R.~S., \& {Backer}, D.~C. 1990, \apj, 361, 300,
  \dodoi{10.1086/169195}

\bibitem[{Gelman {et~al.}(2013)Gelman, Carlin, Stern, Dunson, Vehtari, \&
  Rubin}]{gelman2013bayesian}
Gelman, A., Carlin, J., Stern, H., {et~al.} 2013, Bayesian Data Analysis, Third
  Edition, Chapman \& Hall/CRC Texts in Statistical Science (Taylor \&
  Francis).
\newblock \url{https://books.google.com/books?id=ZXL6AQAAQBAJ}

\bibitem[{Gelman \& Hill(2007)}]{gelman:2007}
Gelman, A., \& Hill, J. 2007, Data analysis using regression and
  multilevel/hierarchical models, Vol. Analytical methods for social research
  (New York: Cambridge University Press), xxii, 625 p

\bibitem[{Godsill(2001)}]{g01}
Godsill, S.~J. 2001, Journal of Computational and Graphical Statistics, 10,
  230.
\newblock \url{http://www.jstor.org/stable/1391010}

\bibitem[{{Hazboun} {et~al.}(2018){Hazboun}, {Mingarelli}, \&
  {Lee}}]{hazboun:2018mdc}
{Hazboun}, J.~S., {Mingarelli}, C. M.~F., \& {Lee}, K. 2018, arXiv e-prints,
  arXiv:1810.10527.
\newblock \doarXiv{1810.10527}

\bibitem[{{Hazboun} {et~al.}(2020){Hazboun}, {Simon}, {Taylor}, {Lam},
  {Vigeland}, {Islo}, {Key}, {Arzoumanian}, {Baker}, {Brazier}, {Brook},
  {Burke-Spolaor}, {Chatterjee}, {Cordes}, {Cornish}, {Crawford}, {Crowter},
  {Cromartie}, {DeCesar}, {Demorest}, {Dolch}, {Ellis}, {Ferdman}, {Ferrara},
  {Fonseca}, {Garver-Daniels}, {Gentile}, {Good}, {Holgado}, {Huerta},
  {Jennings}, {Jones}, {Jones}, {Kaiser}, {Kaplan}, {Kelley}, {Lazio}, {Levin},
  {Lommen}, {Lorimer}, {Luo}, {Lynch}, {Madison}, {McLaughlin}, {McWilliams},
  {Mingarelli}, {Ng}, {Nice}, {Pennucci}, {Pol}, {Ransom}, {Ray}, {Siemens},
  {Spiewak}, {Stairs}, {Stinebring}, {Stovall}, {Swiggum}, {Turner},
  {Vallisneri}, {van Haasteren}, {Witt}, \& {Zhu}}]{hazboun:2020slice}
{Hazboun}, J.~S., {Simon}, J., {Taylor}, S.~R., {et~al.} 2020, \apj, 890, 108,
  \dodoi{10.3847/1538-4357/ab68db}

\bibitem[{Hee {et~al.}(2015)Hee, Handley, Hobson, \& Lasenby}]{hee15}
Hee, S., Handley, W.~J., Hobson, M.~P., \& Lasenby, A.~N. 2015, Monthly Notices
  of the Royal Astronomical Society, 455, 2461, \dodoi{10.1093/mnras/stv2217}

\bibitem[{{Heinrich} {et~al.}(2004){Heinrich}, {Blocker}, {Conway},
  {Demortier}, {Lyons}, {Punzi}, \& {Sinervo}}]{heinrich:2004}
{Heinrich}, J., {Blocker}, C., {Conway}, J., {et~al.} 2004, arXiv e-prints,
  physics/0409129.
\newblock \doarXiv{physics/0409129}

\bibitem[{{Hellings} \& {Downs}(1983)}]{hd83}
{Hellings}, R.~W., \& {Downs}, G.~S. 1983, 265, L39, \dodoi{10.1086/183954}

\bibitem[{{Hobbs} \& {Edwards}(2012)}]{tempo2}
{Hobbs}, G., \& {Edwards}, R. 2012, {Tempo2: Pulsar Timing Package}.
\newblock \doeprint{1210.015}

\bibitem[{{Holgado} {et~al.}(2018){Holgado}, {Sesana}, {Sandrinelli}, {Covino},
  {Treves}, {Liu}, \& {Ricker}}]{hss+18}
{Holgado}, A.~M., {Sesana}, A., {Sandrinelli}, A., {et~al.} 2018, \mnras, 481,
  L74, \dodoi{10.1093/mnrasl/sly158}

\bibitem[{{Jaffe} \& {Backer}(2003)}]{jb03}
{Jaffe}, A.~H., \& {Backer}, D.~C. 2003, Astrophysical Journal, 583, 616,
  \dodoi{10.1086/345443}

\bibitem[{Kass \& Wasserman(1996)}]{kass+1996}
Kass, R.~E., \& Wasserman, L. 1996, Journal of the American Statistical
  Association, 91, 1343, \dodoi{10.1080/01621459.1996.10477003}

\bibitem[{{Lentati} {et~al.}(2015){Lentati}, {Taylor}, {Mingarelli}, {Sesana},
  {Sanidas}, {Vecchio}, {Caballero}, {Lee}, {van Haasteren}, {Babak}, {Bassa},
  {Brem}, {Burgay}, {Champion}, {Cognard}, {Desvignes}, {Gair}, {Guillemot},
  {Hessels}, {Janssen}, {Karuppusamy}, {Kramer}, {Lassus}, {Lazarus}, {Liu},
  {Os{\l}owski}, {Perrodin}, {Petiteau}, {Possenti}, {Purver}, {Rosado},
  {Smits}, {Stappers}, {Theureau}, {Tiburzi}, \& {Verbiest}}]{lentati2015}
{Lentati}, L., {Taylor}, S.~R., {Mingarelli}, C.~M.~F., {et~al.} 2015, \mnras,
  453, 2576, \dodoi{10.1093/mnras/stv1538}

\bibitem[{Lentati {et~al.}(2016)}]{Lentati:2016ygu}
Lentati, L., {et~al.} 2016, Mon. Not. Roy. Astron. Soc., 458, 2161,
  \dodoi{10.1093/mnras/stw395}

\bibitem[{{Phinney}(2001)}]{p01}
{Phinney}, E.~S. 2001, ArXiv Astrophysics e-prints

\bibitem[{{Rosado} {et~al.}(2015){Rosado}, {Sesana}, \& {Gair}}]{rsg2015}
{Rosado}, P.~A., {Sesana}, A., \& {Gair}, J. 2015, \mnras, 451, 2417,
  \dodoi{10.1093/mnras/stv1098}

\bibitem[{{Sazhin}(1978)}]{saz78}
{Sazhin}, M.~V. 1978, \sovast, 22, 36

\bibitem[{{Sesana} {et~al.}(2018){Sesana}, {Haiman}, {Kocsis}, \&
  {Kelley}}]{shk+18}
{Sesana}, A., {Haiman}, Z., {Kocsis}, B., \& {Kelley}, L.~Z. 2018, \apj, 856,
  42, \dodoi{10.3847/1538-4357/aaad0f}

\bibitem[{{Shannon} {et~al.}(2015){Shannon}, {Ravi}, {Lentati}, {Lasky},
  {Hobbs}, {Kerr}, {Manchester}, {Coles}, {Levin}, {Bailes}, {Bhat},
  {Burke-Spolaor}, {Dai}, {Keith}, {Os{\l}owski}, {Reardon}, {van Straten},
  {Toomey}, {Wang}, {Wen}, {Wyithe}, \& {Zhu}}]{srl+15}
{Shannon}, R.~M., {Ravi}, V., {Lentati}, L.~T., {et~al.} 2015, Science, 349,
  1522, \dodoi{10.1126/science.aab1910}

\bibitem[{Siemens {et~al.}(2013)Siemens, Ellis, Jenet, \& Romano}]{sejr13}
Siemens, X., Ellis, J., Jenet, F., \& Romano, J.~D. 2013, Classical and Quantum
  Gravity, 30, 224015.
\newblock \url{http://stacks.iop.org/0264-9381/30/i=22/a=224015}

\bibitem[{{Simon} \& {Burke-Spolaor}(2016)}]{sbs16}
{Simon}, J., \& {Burke-Spolaor}, S. 2016, \apj, 826, 11,
  \dodoi{10.3847/0004-637X/826/1/11}

\bibitem[{{Taylor} {et~al.}(2013){Taylor}, {Gair}, \& {Lentati}}]{taylor2013}
{Taylor}, S.~R., {Gair}, J.~R., \& {Lentati}, L. 2013, \prd, 87, 044035,
  \dodoi{10.1103/PhysRevD.87.044035}

\bibitem[{{Taylor} {et~al.}(2017){Taylor}, {Lentati}, {Babak}, {Brem}, {Gair},
  {Sesana}, \& {Vecchio}}]{tlb+17}
{Taylor}, S.~R., {Lentati}, L., {Babak}, S., {et~al.} 2017, \prd, 95, 042002,
  \dodoi{10.1103/PhysRevD.95.042002}

\bibitem[{{Taylor} {et~al.}(2020){Taylor}, {van Haasteren}, \&
  {Sesana}}]{taylor2020}
{Taylor}, S.~R., {van Haasteren}, R., \& {Sesana}, A. 2020, arXiv e-prints,
  arXiv:2006.04810.
\newblock \doarXiv{2006.04810}

\bibitem[{{Tiburzi} {et~al.}(2016){Tiburzi}, {Hobbs}, {Kerr}, {Coles}, {Dai},
  {Manchester}, {Possenti}, {Shannon}, \& {You}}]{thk+2016}
{Tiburzi}, C., {Hobbs}, G., {Kerr}, M., {et~al.} 2016, \mnras, 455, 4339,
  \dodoi{10.1093/mnras/stv2143}

\bibitem[{{Vallisneri}(2020)}]{libstempo}
{Vallisneri}, M. 2020, {libstempo: Python wrapper for Tempo2}.
\newblock \doeprint{2002.017}

\bibitem[{{Vallisneri} {et~al.}(2020){Vallisneri}, {Taylor}, {Simon},
  {Folkner}, {Park}, {Cutler}, {Ellis}, {Lazio}, {Vigeland}, {Aggarwal},
  {Arzoumanian}, {Baker}, {Brazier}, {Brook}, {Burke-Spolaor}, {Chatterjee},
  {Cordes}, {Cornish}, {Crawford}, {Cromartie}, {Crowter}, {DeCesar},
  {Demorest}, {Dolch}, {Ferdman}, {Ferrara}, {Fonseca}, {Garver-Daniels},
  {Gentile}, {Good}, {Hazboun}, {Holgado}, {Huerta}, {Islo}, {Jennings},
  {Jones}, {Jones}, {Kaplan}, {Kelley}, {Key}, {Lam}, {Levin}, {Lorimer},
  {Luo}, {Lynch}, {Madison}, {McLaughlin}, {McWilliams}, {Mingarelli}, {Ng},
  {Nice}, {Pennucci}, {Pol}, {Ransom}, {Ray}, {Siemens}, {Spiewak}, {Stairs},
  {Stinebring}, {Stovall}, {Swiggum}, {van Haasteren}, {Witt}, \&
  {Zhu}}]{vts+20}
{Vallisneri}, M., {Taylor}, S.~R., {Simon}, J., {et~al.} 2020, \apj, 893, 112,
  \dodoi{10.3847/1538-4357/ab7b67}

\bibitem[{van Haasteren \& Levin(2013)}]{vanHaasteren:2012hj}
van Haasteren, R., \& Levin, Y. 2013, Mon. Not. Roy. Astron. Soc., 428, 1147,
  \dodoi{10.1093/mnras/sts097}

\bibitem[{{van Haasteren} \& {Vallisneri}(2014)}]{vHv:2014}
{van Haasteren}, R., \& {Vallisneri}, M. 2014, \prd, 90, 104012,
  \dodoi{10.1103/PhysRevD.90.104012}

\bibitem[{{Vigeland} {et~al.}(2018){Vigeland}, {Islo}, {Taylor}, \&
  {Ellis}}]{vite18}
{Vigeland}, S.~J., {Islo}, K., {Taylor}, S.~R., \& {Ellis}, J.~A. 2018, \prd,
  98, 044003, \dodoi{10.1103/PhysRevD.98.044003}

\bibitem[{Williams \& Rasmussen(2006)}]{rw06}
Williams, C.~K., \& Rasmussen, C.~E. 2006, the MIT Press, 2, 4

\bibitem[{{Yardley} {et~al.}(2011){Yardley}, {Coles}, {Hobbs}, {Verbiest},
  {Manchester}, {van Straten}, {Jenet}, {Bailes}, {Bhat}, {Burke-Spolaor},
  {Champion}, {Hotan}, {Oslowski}, {Reynolds}, \& {Sarkissian}}]{yardley2011}
{Yardley}, D.~R.~B., {Coles}, W.~A., {Hobbs}, G.~B., {et~al.} 2011, \mnras,
  414, 1777, \dodoi{10.1111/j.1365-2966.2011.18517.x}

\end{thebibliography}
\bibliographystyle{aasjournal}

\end{document}